\documentclass[aps,pra,10pt,tightenlines,longbibliography,notitlepage,amsmath,twocolumn,floatfix,superscriptaddress]{revtex4-1}
\usepackage[dvipsnames]{xcolor}
\usepackage[colorlinks=true,bookmarks=false,linkcolor=NavyBlue,urlcolor=NavyBlue,citecolor=NavyBlue,breaklinks]{hyperref}
\usepackage{amsmath,amsfonts,amssymb,amsthm}
\usepackage{mathtools}
\usepackage{tabularx}
\usepackage{graphicx}
\usepackage[italicdiff]{physics}
\usepackage{times}
\usepackage{enumitem}
\usepackage{setspace}

\newcommand{\intall}{\int_{-\infty}^{\infty}}

\newcommand{\bk}[1]{\left(#1\right)}
\newcommand{\Bk}[1]{\left[#1\right]}
\newcommand{\BK}[1]{\left\{#1\right\}}
\newcommand{\vac}{\textrm{vac}}

\DeclareMathOperator{\expect}{\mathbb E}

\DeclareMathOperator{\MSE}{MSE}

\begin{document}

\title{Quantum noise spectroscopy as an incoherent imaging problem}

\author{Mankei Tsang}
\email{mankei@nus.edu.sg}
\homepage{https://blog.nus.edu.sg/mankei/}
\affiliation{Department of Electrical and Computer Engineering,
  National University of Singapore, 4 Engineering Drive 3, Singapore
  117583}

\affiliation{Department of Physics, National University of Singapore,
  2 Science Drive 3, Singapore 117551}

\date{\today}


\begin{abstract}
  I point out the mathematical correspondence between an incoherent
  imaging model proposed by my group in the study of quantum-inspired
  superresolution [Tsang, Nair, and Lu, Physical Review X \textbf{6},
  031033 (2016)] and a noise spectroscopy model also proposed by us
  [Tsang and Nair, Physical Review A \textbf{86}, 042115 (2012); Ng
  \emph{et al.}, Physical Review A \textbf{93}, 042121 (2016)]. Both
  can be regarded as random displacement models, where the probability
  measure for the random displacement depends on unknown
  parameters. The spatial-mode demultiplexing (SPADE) method proposed
  for imaging is analogous to the spectral photon counting method
  proposed in Ng \emph{et al.}~(2016) for optical phase noise
  spectroscopy---Both methods are discrete-variable measurements that
  are superior to direct displacement measurements (direct imaging or
  homodyne detection) and can achieve the respective quantum limits.
  Inspired by SPADE, I propose a modification of spectral photon
  counting when the input field is squeezed---simply unsqueeze the
  output field before spectral photon counting. I show that this
  method is quantum-optimal and far superior to homodyne detection for
  both parameter estimation and detection, thus solving the open
  problems in Tsang and Nair (2012) and Ng \emph{et al.}~(2016).
\end{abstract}

\maketitle

\section{Introduction}
\begin{figure}[htbp!]
\centerline{\includegraphics[width=0.48\textwidth]{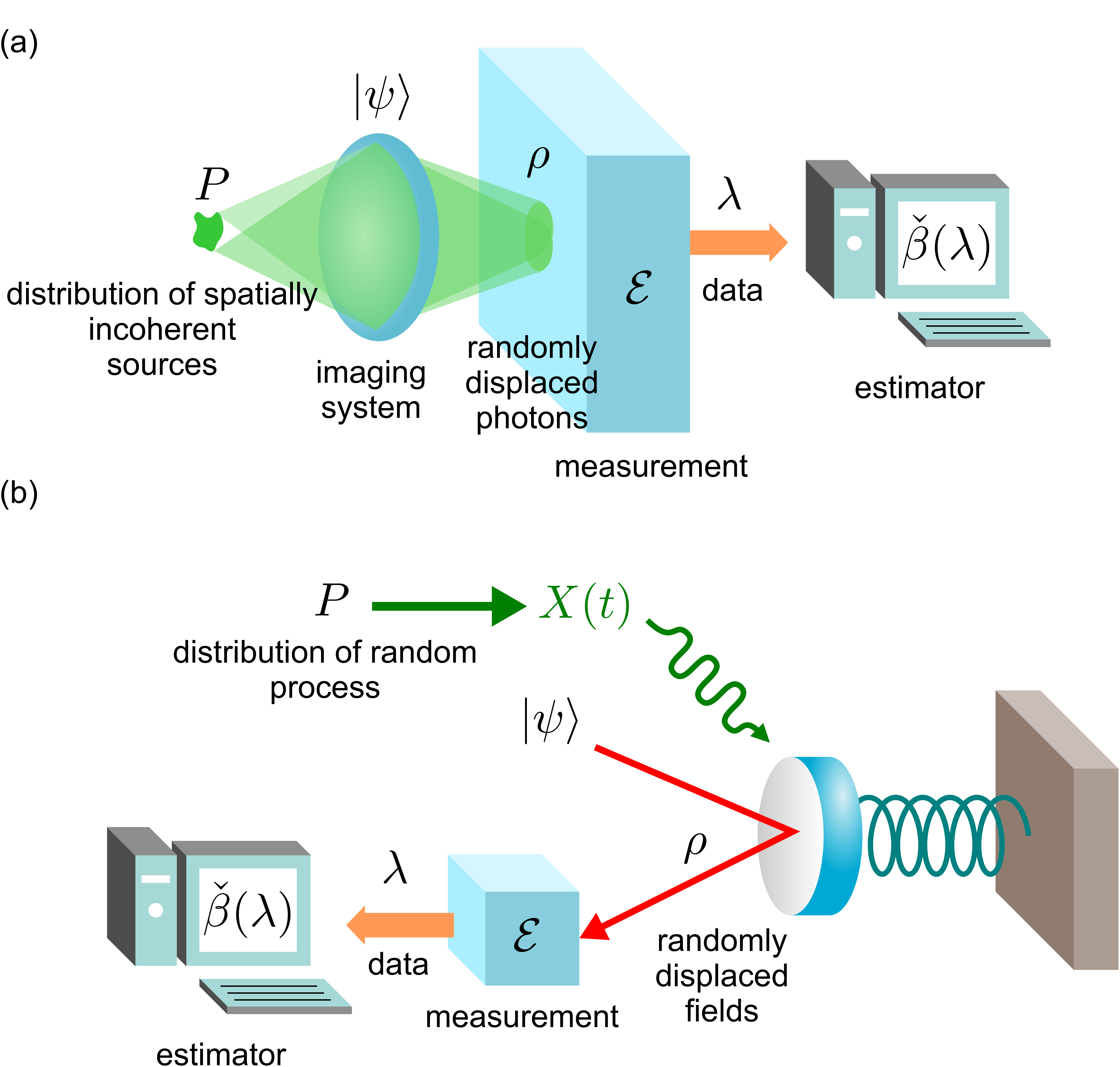}}
\caption{\label{imaging_random_displacement}(a) A schematic of an
    incoherent imaging system, where the quantum state $\rho$ of each
    image-plane photon can be modeled as a randomly displaced object,
    with the distribution of the incoherent sources determining the
    probability measure $P$ for the displacement and the point-spread
    function of the imaging system determining the initial state
    $\ket{\psi}$ of the photon.  (b) A schematic of an optomechanical
    sensor, where the quantum state $\rho$ of the optical fields
    before the measurement can also be modeled as a randomly displaced
    object, with a probability measure $P$ governing the displacement
    and the initial state of the optical fields determining
    $\ket{\psi}$. In both cases, a measurement is modeled by a
    positive operator-valued measure (POVM) $\mathcal E$, and an
    estimator $\check\beta(\lambda)$ of a parameter of $P$ can be
    constructed from the measurement outcome $\lambda$.}
\end{figure}

Optical telescopes and gravitational-wave detectors are two of the
most important technologies in modern physics and astronomy. This
paper studies a remarkable connection between them from the
perspective of quantum metrology. The key insight is that the photons
from incoherent sources received by a telescope and an optomechanical
system under a stochastic gravitational-wave background can both be
modeled as quantum systems under random displacements, as depicted in
Fig.~\ref{imaging_random_displacement}.  In both imaging and the
sensing of stochastic gravitational-wave backgrounds, measurements are
performed to estimate the probabilistic properties of the
displacements, and the measurements for both problems turn out to
share significant similarities in a statistical sense. My group has
studied both problems \cite{tnl,tsang19a,tsang_nair,ng16}, but the
connection has hitherto not been elaborated. Inspired by the
connection, here I use the insights gained from our study of
incoherent imaging to devise an optimal measurement for an optical
random displacement model with squeezed light, thus solving the open
problems in Refs.~\cite{tsang_nair,ng16}. The optimal measurement is
far superior to the standard homodyne detection in the same way
quantum-inspired imaging methods can beat direct imaging.  Beyond
imaging, optomechanics, and gravitational-wave detection, the random
displacement model is also relevant to magnetometers under fluctuating
magnetic fields \cite{budker07} and microwave cavities driven by
hypothetical dark-matter axions \cite{backes21}, so the insights and
results here should have wider implications.

\section{Models}
Consider first the incoherent imaging system depicted in
  Fig.~\ref{imaging_random_displacement}(a). The one-photon density
operator $\rho$ on the image plane can be modeled as
\cite{tnl,tsang19a}
\begin{align}
\rho &= \int dP\ U_{X} \ket{\psi}\bra{\psi} U_{X}^\dagger,
\label{rho}
\\
U_X &= \prod_{m=1}^M \exp(-i k_m X_m),
\label{UX_imaging}
\end{align}
where $M$ is the dimension of the object and image planes,
$\ket{\psi}$, an element of the Hilbert space
$\mathcal H = \mathcal H_1 \otimes \dots \otimes \mathcal H_M$, models
the diffraction-limited point-spread function of the imaging system,
$k_m$ is a momentum operator on $\mathcal H_m$, $U_X$ is a unitary
operator that models the photon displacement on the image plane due to
a point source, and $X$ is a real classical $M$-dimensional random
vector under the probability measure $P$, which models the object
intensity function. Mathematically, Eq.~(\ref{rho}) is a Bochner
integral; both $dP$ and $X$ in Eq.~(\ref{rho}) depend implicitly on
$x \in S$ in terms of a probability space $(S,\Sigma,P)$
\cite{holevo11}.

The imaging problem can be framed as a quantum detection or estimation
problem \cite{tsang19a,helstrom}, where $P$ belongs to a family of
probability measures $\{P_\theta:\theta \in \Theta\}$ parametrized by
a parameter $\theta$ in some parameter space $\Theta$ and a parameter
of interest $\beta(\theta)$ is to be estimated via measurements of the
optical fields. Studies in the area of quantum-inspired
superresolution have shown that spatial-mode demultiplexing (SPADE)
can offer a far superior performance over direct imaging and achieve
the quantum limits in the resolution of two point sources
\cite{tnl,lu18}, object-size estimation \cite{tsang17,dutton19}, and
moment estimation
\cite{tsang17,tsang18a,tsang19,zhou19,tsang19b,tsang21a,tsang22}.  For
the uninitiated, Appendices~\ref{app_est} and \ref{app_qisr} offer a
brief review of these results.

The incoherent imaging model turns out to be mathematically similar to
a noise spectroscopy model also proposed by my group in
Refs.~\cite{tsang_nair,ng16}. The main difference is in the dimension
$M$: imaging problems usually assume that $M$ is one or two, whereas
Refs.~\cite{tsang_nair,ng16} assume that it is infinite.  In the noise
spectroscopy problem, $\rho$ is the state of a quantum dynamical
system coupled to quantum fields, $\ket{\psi}$ is an element of an
infinite-dimensional Hilbert space $\mathcal H$ that models the input
state of the total system, $X(t)$ is a real classical random process
with respect to a time variable $t$ that generalizes the $m$ in
Eq.~(\ref{UX_imaging}), and the unitary is
\begin{align}
U_X &= \mathcal T \exp[-i \int_{0}^{T} dt k(t) X(t)],
\label{UX}
\end{align}
where $\mathcal T$ denotes time ordering, $T$ is the total observation
time, and $k(t)$ is a Hermitian operator on $\mathcal H$ in an
interaction picture \cite{tsang_open}.  $\ket{\psi}$ and $k(t)$ are
assumed to be independent of $X$. Any sequential measurements
concurrent with the displacement can be modeled as a final measurement
via the principle of deferred measurement \cite{twc,nielsen}.
Examples include an optical field under a random displacement or phase
modulation, an optomechanical system under a stochastic force
\cite{aspelmeyer14,nimmrichter}, a gravitational-wave detector under a
stochastic background \cite{christensen19}, a spin ensemble under a
stochastic magnetic field \cite{budker07}, and a microwave cavity
driven by dark-matter axions \cite{backes21}.
Figure~\ref{imaging_random_displacement}(b) depicts an
  optomechanical system as an example.

References~\cite{tsang_nair,ng16} assume that $X(t)$ is a stationary
zero-mean Gaussian random process, and its power spectral density
$S_X(\omega|\theta)$ depends on the unknown parameter
$\theta$. Reference~\cite{tsang_nair} assumes that $\Theta$ is binary
with $S_X(\omega) = 0$ for one of the hypotheses, such that the
problem of interest is the detection of a random displacement, while
Ref.~\cite{ng16} assumes that $\Theta$ is a multidimensional Euclidean
space, such that the problem is spectrum-parameter estimation. In
other words, Refs.~\cite{tsang_nair,ng16} assume parametric models for
the probability measure $P$, in the same way parametric models for $P$
are assumed for incoherent imaging.

\section{\label{sec_est}Spectrum-parameter estimation}
The power spectral density, being a second-order statistic, is
analogous to the second-order object moments in the context of
imaging. Since SPADE can enhance the estimation of second-order
moments
\cite{tsang17,tsang18a,tsang19,zhou19,tsang19b,tsang21a,tsang22}, it
is natural to ask if a similar enhancement can be found for noise
spectroscopy. The answer is yes---Ref.~\cite{ng16} considers an
optical field under weak and random phase modulation and finds that
spectral photon counting, a discrete-variable measurement analogous to
SPADE, can be far superior to homodyne detection, a
continuous-variable measurement analogous to direct imaging, when the
input state $\ket{\psi}$ is a coherent state. Spectral photon counting
is quantum-optimal and enjoys significant superiority over homodyne
detection in the regime of low signal-to-noise ratios, just as SPADE
is quantum-optimal and superior in the regime of subdiffraction object
sizes for imaging.

In the following, I adopt a level of mathematical rigor typical of the
physics and engineering literature \cite{gardiner_zoller,shapiro09} to
arrive at results quickly, following Refs.~\cite{tsang_nair,ng16}. To
derive a quantum limit to noise spectroscopy, Ref.~\cite{ng16} makes
the following assumptions:
\begin{enumerate}[label=(A\arabic*)]
\item $X(t)$ is a zero-mean Gaussian process.

\item The processes $X(t)$ and
\begin{align}
\Delta k(t) \equiv k(t) - \bra\psi k(t)\ket\psi
\end{align}
are stationary in the wide sense
\cite{vantrees,shumway_stoffer,gardiner_zoller}, such that
\begin{align}
C_X(\tau|\theta) &\equiv \expect_\theta[X(t)X(t+\tau)],
\\
C_k(\tau) &\equiv \bra{\psi}\Delta k(t)\circ \Delta k(t+\tau)\ket{\psi}
\end{align}
are independent of $t$.

($\expect_\theta$ denotes the expectation with respect to $P_\theta$
and $A\circ B \equiv (AB+BA)/2$ denotes the Jordan product.)

\item The observation time $T$ is long enough to justify certain
  approximations regarding stationary processes
  \cite{vantrees,shumway_stoffer}.
  
\end{enumerate}
Such assumptions are common in statistics
\cite{vantrees,shumway_stoffer} and have the virtue of giving simple
closed-form results for the infinite-dimensional model. Assuming also
that $\theta$ is a real scalar for simplicity, a quantum limit to the
Fisher information $J$ for any measurement is \cite{ng16}
\begin{align}
J &\le K \le \tilde K \to 
T \intall\frac{d\omega}{2\pi}\frac{(\partial \ln S_X)^2}{2 + 1/(S_k S_X)} ,
\label{ec}
\\
S_X(\omega|\theta)
&\equiv 
\intall d\tau C_X(\tau|\theta) \exp(i\omega\tau),
\\
S_k(\omega) &\equiv 
\intall d\tau C_k(\tau) \exp(i\omega \tau),
\label{Sk}
\end{align}
where $K$ is the Helstrom information in terms of $\rho$ as a function
of $\theta$ \cite{helstrom,hayashi}, $\tilde K$ is a bound derived in
Ref.~\cite{ng16} using the extended convexity of $K$ \cite{alipour},
$\partial \equiv \pdv*{}{\theta}$, and $\to$ denotes the long-time
limit. These information quantities determine the fundamental
  limits to the estimation of a parameter $\theta$ of the power
  spectral density $S_X(\omega|\theta)$ of the ``noise'' $X(t)$ in
  noise spectroscopy. For the uninitiated, Appendices~\ref{app_est}
  and \ref{app_ec} give a brief review of the basic concepts in
  quantum estimation theory.

  Note that Eq.~(\ref{ec}) is applicable to scenarios where the probe
  is initially entangled with an ancilla, as $\ket{\psi}$ can model
  the initial state of the probe plus the ancilla in a larger Hilbert
  space and $U_X$ can be an operator on the probe subspace only.

At the time of Ref.~\cite{ng16}, we were unable to find a
quantum-optimal measurement when the input state is not a coherent
state, but the correspondence with incoherent imaging offers a new
insight.  We know that SPADE can remain superior and optimal as long
as its basis is adapted to $\ket{\psi}$
\cite{tnl,rehacek17,tsang18a,tsang21a,lu18}. This fact suggests that a
discrete-variable measurement is still optimal for noise spectroscopy
with a nonclassical state, as long as the measurement basis is adapted
to the input state $\ket{\psi}$. If $\ket{\psi}$ is a squeezed state,
it still has a Gaussian wavefunction and is analogous to a Gaussian
point-spread function in imaging. The imaging correspondence then
suggests that an optimal basis adapted to $\ket{\psi}$ is simply a
squeezed version of an optimal basis adapted to the vacuum. A
measurement in that basis can be implemented by unsqueezing the output
field, analogous to an image magnification, before spectral photon
counting.

I now show the optimality of the unsqueezing and spectral photon
counting (USPC) method in detail. Let
\begin{align}
k(t) &= A^\dagger(t) A(t),
&
\qty[A(t),A^\dagger(t')] &= \delta(t-t'),
\end{align}
where $A(t)$ is the annihilation operator for the slowly varying
envelope of an optical field with carrier frequency $\Omega$
\cite{gardiner_zoller,shapiro09} and $k(t)$ is the photon-flux
operator. $X$ is then a phase modulation on the optical field. Since
$k(t)$ commutes with itself at different times, the time ordering in
Eq.~(\ref{UX}) is redundant. Assume also
\begin{align}
\ket{\psi} &= D(\alpha) V\ket{\vac},
\label{sq_vac}
\end{align}
where $\ket{\vac}$ is the vacuum state, $V$ is a unitary operator that
models the squeezing, and $D(\alpha)$ is the displacement operator
that gives a constant mean field
$\bra{\psi} A(t)\ket{\psi} = \alpha$. $|\alpha|^2$ is the mean
photon flux. With a high $|\alpha|$ and weak phase modulation,
$D^\dagger k(t) D$ can be linearized as an intensity quadrature
operator
\begin{align}
D^\dagger k(t) D &\approx |\alpha|^2 + \kappa(t),
&
\kappa(t) &\equiv \alpha A^\dagger(t) +\alpha^* A(t),
\label{linearize}
\end{align}
and $D^\dagger U_X D$ becomes a displacement operator.  This
  linearization turns the phase modulation into a displacement. The
initial squeezing $V$ should squeeze the orthogonal phase quadrature
\begin{align}
\eta(t) &\equiv \frac{1}{2i|\alpha|^2}
\qty[\alpha A^\dagger(t) -\alpha^* A(t)]
\end{align} 
and antisqueeze the intensity quadrature, such that
\begin{align}
V^\dagger \eta V &= h * \eta,
&
V^\dagger  \kappa V &= g * \kappa,
\end{align}
where $h *\eta \equiv \intall dt' h(t-t') \eta(t')$ denotes the
convolution and the real Green functions $h(t)$ and $g(t)$ model the
squeezing and the antisqueezing, respectively
\cite{gardiner_zoller}. Their Fourier transforms are related by
\begin{align}
|\tilde h(\omega) \tilde g(\omega)| &= 1,
\end{align}
where
\begin{align}
\tilde g(\omega) &\equiv \intall dt g(t)\exp(i\omega t),
\end{align}
and $\tilde h(\omega)$ is defined similarly.

\begin{figure}[htbp!]
\centerline{\includegraphics[width=0.48\textwidth]{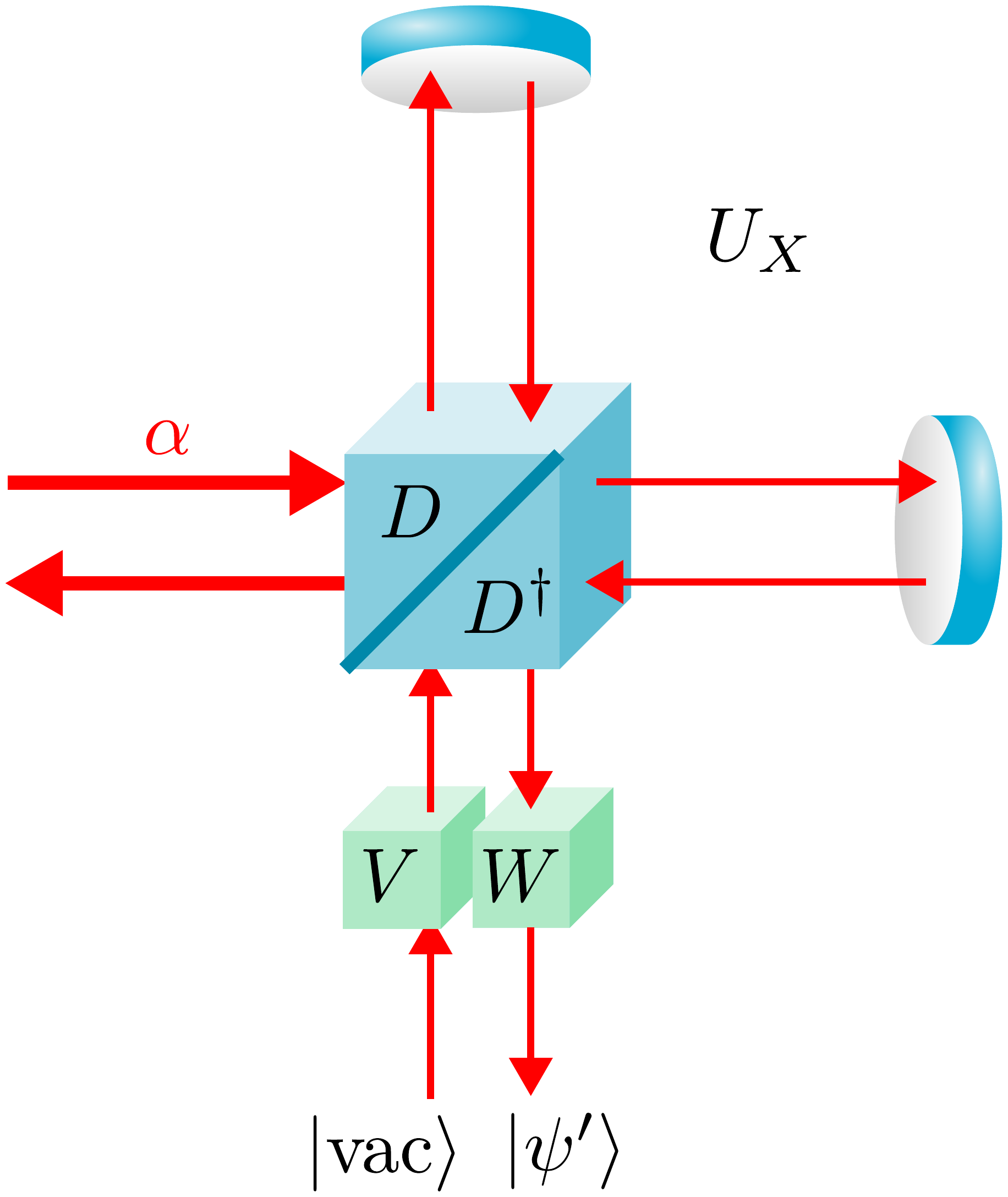}}
\caption{\label{michelson}A Michelson implementation of the model
    given by Eqs.~(\ref{sq_vac})--(\ref{mean_field}).}
\end{figure}

After $U_X$, suppose that the mean field is nulled by
$D^\dagger(\alpha)$ and then the squeezing is undone by a unitary $W$,
which is the same as $V$ except that a negative sign is introduced to
the parametric-amplifier Hamiltonian. In other words, the
  experimental setup for the unsqueezing can be the same as that for
  $V$, except that the phase of the pump beam should be shifted by
  $\pi$ if the parametric amplifier is implemented by three-wave
  mixing.  The effect of $W$ on the quadratures can be modeled as
\begin{align}
W^\dagger \eta W &= g * \eta,
&
W^\dagger \kappa W &= h * \kappa.
\label{unsqueeze}
\end{align}
Note that $W$ is not $V^\dagger$, as the Green functions would become
anticausal and thus unphysical if $W$ were $V^\dagger$. 
Conditioned
on $X$, the output state
\begin{align}
\ket{\psi'} = W D^\dagger U_X D V \ket{\vac}
\label{psi_out}
\end{align}
is a coherent state with mean field
\begin{align}
\alpha'(t) &\equiv  \bra{\psi'}A(t)\ket{\psi'} = -i \alpha g * X,
\label{mean_field}
\end{align}
where the displacement in the phase quadrature is amplified by the
unsqueezing. This model is also applicable to the dark port of a
Michelson interferometer \cite{caves81}, where the squeezing $V$ and
the unsqueezing $W$ should be applied to the input and output of the
dark port, respectively, the displacements $D$ and $D^\dagger$ are
naturally implemented by a strong beam at the other input port and the
beam splitter in the interferometer, and $X$ is proportional to the
relative phase between the two arms, as depicted by
  Fig.~\ref{michelson}. Any radiation-pressure-induced noise is
assumed to be negligible or eliminated \cite{klmtv,qnc}.

To facilitate the analysis of the subsequent step of spectral photon
counting, I discretize frequency by assuming that $X(t)$ is given by
the Fourier series
\begin{align}
X(t) &= \frac{1}{\sqrt{T}}\sum_{m=-\infty}^\infty \tilde X_m\exp(-i\omega_m t),
\quad
\omega_m \equiv \frac{2\pi m}{T},
\label{series}
\\
\tilde X_m &\equiv \frac{1}{\sqrt{T}}\int_0^T dt X(t) \exp(i\omega_m t).
\end{align}
Then the mean field of the output coherent state given by
Eq.~(\ref{mean_field}) can be expressed as
\begin{align}
\alpha'(t) &= -\frac{i\alpha}{\sqrt{T}} \sum_{m=-\infty}^\infty
\tilde g(\omega_m) \tilde X_m\exp(-i\omega_m t).
\end{align}
Suppose that a spectrometer disperses the output field in terms of
frequency modes defined by the annihilation operators
\begin{align}
a_m &\equiv \frac{1}{\sqrt{T}}\int_0^T dt A(t) \exp(i\omega_m t),
&
m &\in \mathbb Z,
\end{align}
where $\omega_m$ is a sideband frequency relative to the carrier
$\Omega$ \cite{shapiro98}. Each frequency mode is then in a coherent
state with a displacement given by
\begin{align}
\tilde \alpha_m \equiv \bra{\psi'}a_m\ket{\psi'} = 
\alpha\tilde g(\omega_m) \tilde X_m.
\end{align}
Since $X(t)$ is real,
$\tilde X_{-m} = \tilde X_m^*$. Assume that $\{\tilde X_m: m > 0\}$ are
independent zero-mean complex Gaussian random variables, each with
variance 
\begin{align}
\expect_\theta(|\tilde X_m|^2) &= S_X(\omega_m|\theta).
\end{align}
Assume also that $\tilde X_0$ is a zero-mean real Gaussian random
variable that is independent of the rest. These assumptions allow the
Fourier series given by Eq.~(\ref{series}) to approach any real
stationary zero-mean Gaussian process in the long-time limit
\cite{shumway_stoffer}.  By summing the photon counts at each pair of
sideband frequencies $-\omega_{m}$ and $\omega_m$, one obtains a set
of photon counts that follow the Bose-Einstein distribution
\begin{align}
f_\theta(n) &= \prod_{m> 0}\frac{1}{1+\bar N_m}\qty(\frac{\bar N_m}{1+\bar N_m})^{n_m},
\label{bose}
\\
\bar N_m(\theta) &= 
2\expect_\theta(|\tilde\alpha_m|^2) = 
2 |\alpha\tilde g(\omega_m)|^2 S_X(\omega_m|\theta).
\label{bose_mean}
\end{align}
The $m = 0$ mode has a more complicated photon-count distribution that
need not be considered, as there is a continuum of modes in the
long-time limit and the information provided by one mode should be
negligible.  The Fisher information for USPC is hence
\begin{align}
J_{\textrm{USPC}} &\equiv
\sum_n f_\theta(n) \qty[\partial \ln f_\theta(n)]^2 
\\
&= \sum_{m > 0} \frac{(\partial \ln \bar N_m)^2}{1+1/\bar N_m}
\\
&\to T \intall \frac{d\omega}{2\pi}\frac{(\partial \ln S_X)^2}
{2 + 1/(|\alpha \tilde g|^2 S_X)},
\label{Juspc}
\end{align}
where the long-time limit gives
$\sum_{m> 0} \to T \int_0^\infty d\omega/(2\pi)$ and the integral
$\int_0^\infty d\omega$ for an even integrand is rewritten as the
double-sided integral $\intall d\omega/2$ for easier comparison with
Eq.~(\ref{ec}).

To compare this result with the quantum bound, note that the power
spectral density of $\Delta k(t)$ with respect to
$\ket{\psi} = D V \ket{\vac}$ is the same as that of $\Delta k'(t)$
with respect to $\ket{\vac}$, where
$k'(t) \equiv V^\dagger D^\dagger k(t) D V$, and the antisqueezing of
the intensity quadrature by $V$ leads to
\begin{align}
S_k(\omega) &=  |\alpha\tilde g(\omega)|^2.
\label{Sp}
\end{align}
With this $S_k(\omega)$, the USPC information given by
Eq.~(\ref{Juspc}) matches the quantum bound $\tilde K$ given by
Eq.~(\ref{ec}) and is hence quantum-optimal.

For comparison, the Fisher information for homodyne detection of the
phase quadrature $U_X^\dagger \eta U_X = \eta + X$ is
\cite{ng16,whittle}
\begin{align}
J_{\textrm{hom}} &\to T \intall  \frac{d\omega}{2\pi}
\frac{(\partial \ln S_X)^2}{2 + 4S_\eta/S_X + 2(S_\eta/S_X)^2},
\label{Jhom}
\end{align}
where $\eta$ is assumed to be a stationary zero-mean Gaussian process
with power spectral density $S_\eta$.  For the squeezed $\ket{\psi}$,
\begin{align}
S_\eta(\omega) = \frac{1}{4 S_k(\omega)},
\end{align}
and $S_\eta S_k \ge 1/4$ in general \cite{gardiner_zoller}.  Compared
with the optimal information given by Eqs.~(\ref{ec}) and
(\ref{Juspc}), the homodyne information with a quantum-limited
$S_\eta$ has an extra factor $2(S_\eta/S_X)^2$ in the denominator,
which is significant when the spectral signal-to-noise ratio
(SNR) $S_X/S_\eta$ is low. To see their difference more clearly, assume
\begin{align}
\frac{S_X}{S_\eta} = 4S_k S_X \ll 1
\label{low_snr}
\end{align}
and perform Taylor approximations of Eqs.~(\ref{ec}),
(\ref{Juspc}), and (\ref{Jhom}), which give
\begin{align}
J_{\textrm{USPC}} &\approx \tilde K \approx
T \intall \frac{d\omega}{2\pi} S_k S_X (\partial \ln S_X)^2,
\label{Juspc_low}
\\
J_{\textrm{hom}} &\approx 
8T \intall  \frac{d\omega}{2\pi} (S_kS_X)^2 (\partial \ln S_X)^2.
\label{Jhom_low}
\end{align}
$J_{\textrm{hom}}$ is much lower because of an extra factor of
$8S_kS_X$ in the integrand.

For a simple example, suppose that
$S_X(\omega|\theta) = \theta^2 R(\omega)$, where $\theta$ is the
magnitude of the displacement and $R(\omega)$ is a known spectrum.
In other words, the shape of the noise spectrum is assumed to be
  known, and one is simply interested in estimating the height of
  $\sqrt{S_X(\omega|\theta)}$.  Then
\begin{align}
J_{\textrm{USPC}}  &\approx \tilde K \to 
T \intall  \frac{d\omega}{2\pi}\frac{4}{2\theta^2 + 1/(S_k R)},
\label{Juspc_exa}
\\
J_{\textrm{hom}}
&\to
T \intall  \frac{d\omega}{2\pi}\frac{4}
{2\theta^2 + 1/(S_kR) + 1/[8(\theta S_k R)^2]}.
\end{align}
As $\theta \to 0$, $J_{\textrm{hom}}$ scales quadratically with
$\theta$ and vanishes, while $J_{\textrm{USPC}}$ tends to a positive
constant. These behaviors are analogous to the phenomenon of
Rayleigh's curse for direct imaging and the superiority of SPADE in
two-point resolution and object-size estimation
\cite{tnl,tsang17,dutton19}.

\begin{figure}[htbp!]
\centerline{\includegraphics[width=0.48\textwidth]{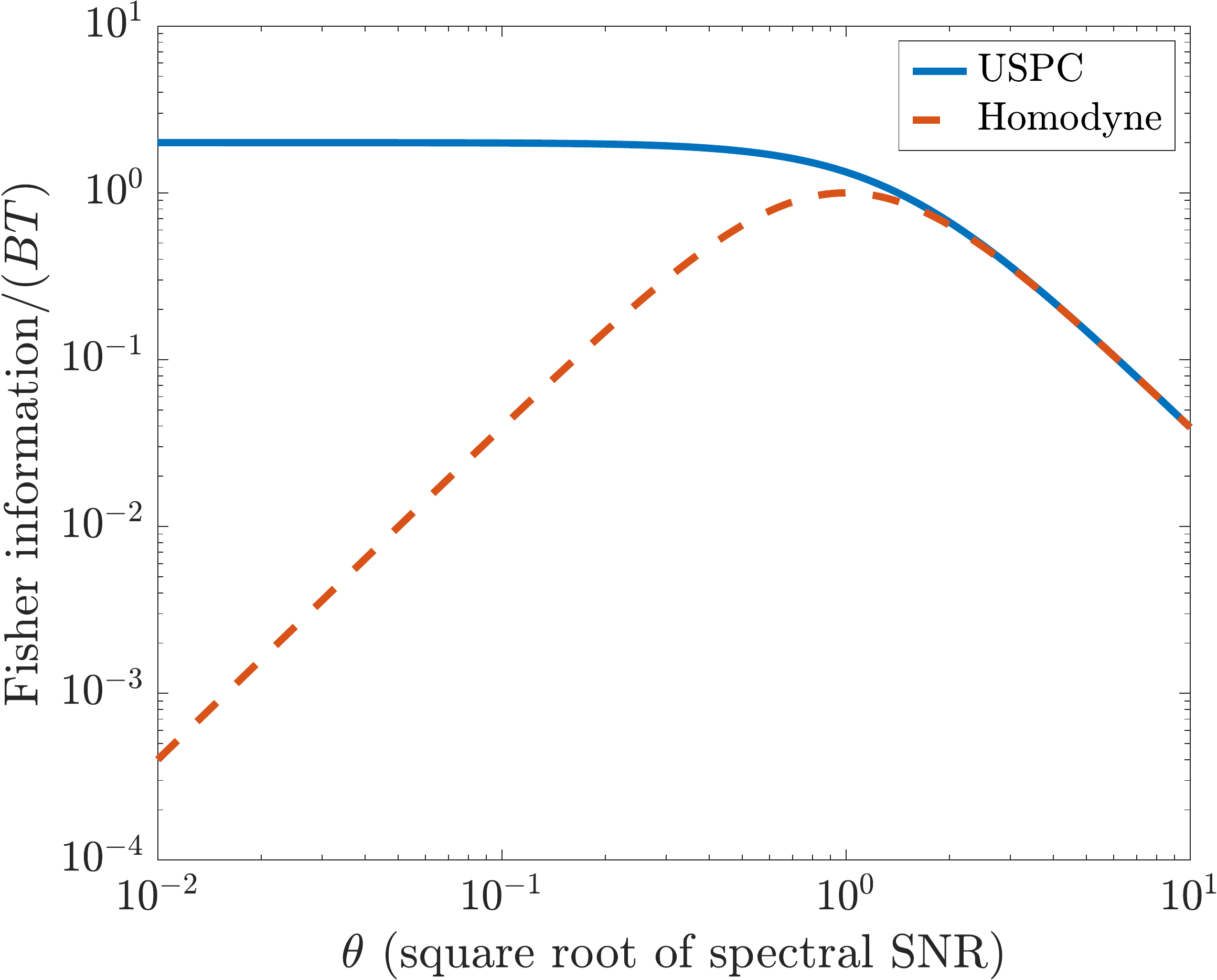}}
\caption{\label{fisher_spec}A comparison of USPC and homodyne
    detection in terms of their Fisher information given by
    Eqs.~(\ref{fisher_simple}) as a function of $\theta$ in log-log
    scale. The Fisher information has been normalized with respect to
    the time-bandwidth product $BT$. $\theta$ has been normalized so
    that $\theta^2$ is the ratio of the displacement spectrum $S_X$ to
    the quantum-limited phase-quadrature spectrum
    $S_\eta = 1/(4 S_k)$.  Both axes are dimensionless.}
\end{figure}

To be even more concrete, suppose that both
  $S_k(\omega) = |\alpha\tilde g(\omega)|^2$ and $R(\omega)$ are flat
  within the band $|\omega| \le 2\pi B$ and $R(\omega) = 0$
  otherwise. Furthermore, assume an $R(0)$ so that $\theta^2$ is the
  spectral SNR. In other words, assume
\begin{align}
S_k(\omega) &= S_k(0) \textrm{ if } |\omega| \le 2\pi B,
\label{Sk_flat}
\\
R(\omega) &= \begin{cases}
S_\eta(0) = 1/[4S_k(0)], & |\omega| \le 2\pi B,
\\
0, &\textrm{otherwise},
\end{cases}
\label{R_flat}
\\
\theta^2 &=\frac{S_X(0|\theta)}{S_\eta(0)} = 
4 S_k(0)S_X(0|\theta).
\end{align}
Then
\begin{align}
J_{\textrm{USPC}}  &\to  \frac{4BT}{\theta^2 + 2},
&
J_{\textrm{hom}} &\to \frac{4BT}{\theta^2 + 2 + 1/\theta^2},
\label{fisher_simple}
\end{align}
which are plotted in log-log scale in Fig.~\ref{fisher_spec}.  Notice
the difference in the scalings for $\theta \ll 1$ and the substantial
widening gap.

In practice, the unknown parameter of the noise spectrum is, of
course, often multidimensional or even the function $S_X$ itself with
no simple parametric model. The results on multiparameter or
semiparametric estimation in imaging offer encouragement that the
superiority of USPC should still persist for those more complicated
problems.

\section{\label{sec_det}Stochastic-displacement detection}
Consider now the detection problem studied in Ref.~\cite{tsang_nair}.
Let $\theta \in \Theta = \{0,1\}$, $P_0$ be the measure that gives the
deterministic $X = 0$ when the displacement is absent, $P_1$ be the
measure for $X$ when the displacement is present, and $\rho_\theta$ be
the quantum state as a function of $\theta$. Since
$\rho_0 = \ket{\psi}\bra{\psi}$ is pure in this problem, the Uhlmann
fidelity is given by
\begin{align}
F &\equiv \trace \sqrt{\sqrt{\rho_0}\rho_1\sqrt{\rho_0}}
= \sqrt{\bra{\psi}\rho_1\ket{\psi}},
\end{align}
while the quantum Chernoff exponent $\zeta$ \cite{audenaert08} is
given by
\begin{align}
\xi &\le \zeta \equiv -\ln \inf_{0\le s \le 1}\trace\qty(\rho_0^{1-s}\rho_1^s)
= - 2\ln F,
\end{align}
where $\xi$ is the classical Chernoff exponent for any measurement.
$F$ and $\zeta$ can be used to set a variety of lower and upper bounds
on the error probabilities under the Bayesian or Neyman-Pearson
criterion; see Appendix~\ref{app_det} for a quick summary of the
  Bayesian theory.

Assuming (A1)--(A3) for $P_1$ and $\ket{\psi}$ and also
\begin{enumerate}
\item[(A4)] $\ket{\psi}$ is a Gaussian state,

\item[(A5)] $k(t)$ is a linear function of bosonic creation and
  annihilation operators, such that $U_X$ is a displacement operator,
\end{enumerate}
we found that the quantum exponent is \cite{tsang_nair}
\begin{align}
\zeta &\to \frac{T}{2} \intall  \frac{d\omega}{2\pi}\ln \qty(1 + 2 S_k S_X).
\label{qexp}
\end{align}
We also considered in Ref.~\cite{tsang_nair} the performances of the
Kennedy receiver and the homodyne detection for the optical model, but
we were unable to find the exact optimal measurement at the time. Here
I solve the open problem by showing that USPC is also optimal for the
detection problem, in analogy with the optimality of SPADE for the
binary-source detection problem \cite{lu18}. Assuming again weak phase
modulation, the USPC distribution given by Eqs.~(\ref{bose}) and
(\ref{bose_mean}), and
\begin{align}
S_X(\omega|0) &= 0,
&
f_0(n) &= \delta_{n0},
&
S_X(\omega|1) &= S_X(\omega),
\end{align}
the Chernoff exponent is
\begin{align}
\xi_{\textrm{USPC}} &\equiv -\ln \inf_{0\le s \le 1}\sum_n \qty[f_0(n)]^{1-s} \qty[f_1(n)]^s
\\
&= \sum_{m> 0} \ln\qty[1+\bar N_m(1)]
\\
&\to \frac{T}{2} \intall  \frac{d\omega}{2\pi}\ln \qty(1 + 2|\alpha\tilde g|^2 S_X).
\label{exp_uspc}
\end{align}
With the $S_k$ given by Eq.~(\ref{Sp}), $\xi_{\textrm{USPC}}$ matches
the quantum limit $\zeta$ given by Eq.~(\ref{qexp}).

For comparison, consider the classical Chernoff exponent for homodyne
detection given by \cite{tsang_nair,shumway_stoffer}
\begin{align}
\xi_{\textrm{hom}} &\to
\sup_{0\le s \le 1}\frac{T}{2} \intall \frac{d\omega}{2\pi}
\ln\qty[\frac{1+(1-s) S_X/S_\eta}{(1 + S_X/S_\eta)^{1-s}}].
\label{exp_hom}
\end{align}
The imaging correspondence suggests that there should be a significant
gap between $\zeta$ and $\xi_{\textrm{hom}}$, although we did not
realize it at the time of Ref.~\cite{tsang_nair}. To demonstrate the
gap now, assume again a low spectral SNR as per
Eq.~(\ref{low_snr}) and perform Taylor approximations of
Eqs.~(\ref{qexp}), (\ref{exp_uspc}), and (\ref{exp_hom}), which give
\begin{align}
\xi_{\textrm{USPC}} &\approx \zeta \approx T \intall \frac{d\omega}{2\pi} S_k S_X ,
\label{exp_uspc_low}
\\
\xi_{\textrm{hom}} &\approx 
\sup_s \frac{s(1-s)T}{4}  \intall \frac{d\omega}{2\pi}
\qty(\frac{S_X}{S_\eta})^2
\\
&= T\intall \frac{d\omega}{2\pi} \qty(S_k S_X)^2.
\label{exp_hom_low}
\end{align}
The optimal exponent is linear with respect to $S_X$, whereas the
homodyne exponent is only quadratic. These scalings are analogous to
the scalings of the optimal exponent and the direct-imaging exponent
with respect to the source separation in the binary-source detection
problem \cite{lu18}.

\begin{figure}[htbp!]
\centerline{\includegraphics[width=0.48\textwidth]{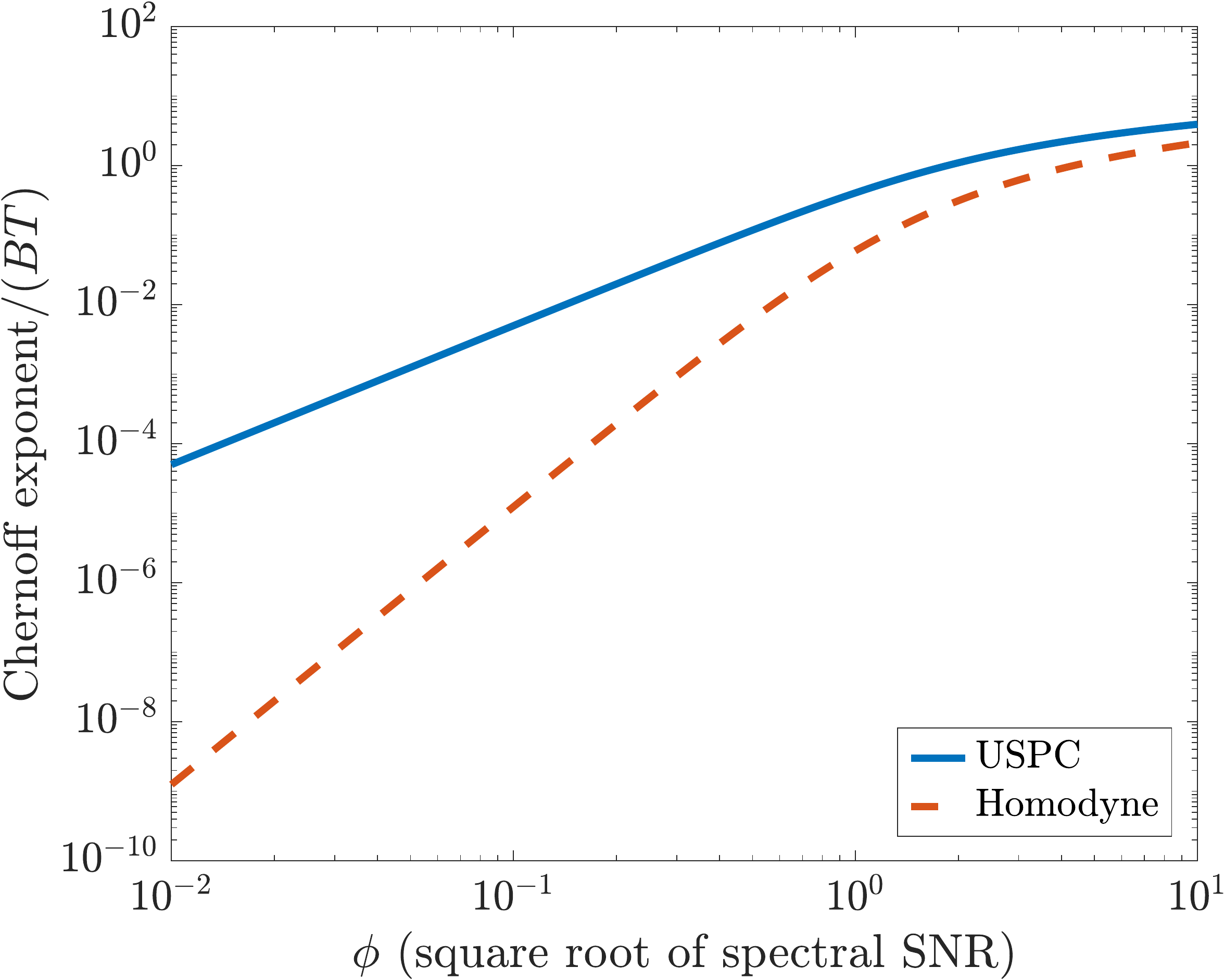}}
\caption{\label{chernoff_spec}A comparison of USPC and homodyne
    detection in terms of their Chernoff exponents given by
    Eqs.~(\ref{exp_uspc}) and (\ref{exp_hom}), assuming the flat
    spectra given by Eqs.~(\ref{Sk_flat}) and (\ref{R_flat}).  The
    horizontal axis $\phi$ is the square root of the spectral SNR
    defined by Eq.~(\ref{phi}). The Chernoff exponent has been
    normalized with respect to the time-bandwidth product $BT$. The
    plot is in log-log scale and both axes are dimensionless.}
\end{figure}

For a more concrete example, assume the flat spectra given by
  Eqs.~(\ref{Sk_flat}) and (\ref{R_flat}) and define
\begin{align}
S_X(\omega) &= \phi^2 R(\omega),
\label{phi}
\end{align}
so that the spectral SNR is now $\phi^2$, and $\phi$ plays the same
role as $\theta$ in Fig.~\ref{fisher_spec}. Figure~\ref{chernoff_spec}
plots the resulting Chernoff exponents given by Eqs.~(\ref{exp_uspc})
and (\ref{exp_hom}) against $\phi$ in log-log scale.

It is possible to study the error probabilities of the detection
problem more precisely under the Neyman-Pearson criterion
\cite{tsang_nair,hayashi,huang21,zanforlin22}, although the insights
offered by such calculations should not deviate much from the ones
reported here.

\section{\label{sec_discuss}Discussion}
Since homodyne detection is the current standard measurement method in
gravitational-wave detection \cite{danilishin19}, the superior
scalings of the USPC information quantities indicated by
Eqs.~(\ref{Juspc_low}), (\ref{Jhom_low}), (\ref{exp_uspc_low}), and
(\ref{exp_hom_low}) are important discoveries. They suggest that USPC
can substantially enhance the detection and spectroscopy of stochastic
gravitational-wave backgrounds when the spectral SNR is low, in the
same way SPADE can enhance incoherent imaging. Considering that
squeezed light is now being used in gravitational-wave detectors
\cite{tse19}, the unsqueezing step proposed here is important, as it
optimizes the measurement for squeezed light beyond the coherent-state
case considered in Refs.~\cite{tsang_nair,ng16} and allows the full
potential of quantum-enhanced interferometry to be realized for noise
spectroscopy.

A potential practical issue with the proposal is its assumption of
quantum-limited squeezing and unsqueezing in both quadratures. Optical
squeezers in current technology often introduce excess noise in the
antisqueezed quadrature, which has little impact on the homodyne
detection of the squeezed quadrature but may add significant noise to
the photon counting step here. With two squeezers in the proposed
setup, the issue of excess noise is even worse. In view of the amazing
achievements of the experimentalists in LIGO and squeezing, however,
one should never underestimate their skills, and the superiority of
USPC should motivate the current and future generations to reach even
greater heights in squeezing technology in order to achieve the
promised improvement.

The correspondence between the incoherent imaging model and the random
displacement model is used implicitly in Section~6 of
Ref.~\cite{tsang17} and briefly mentioned in Ref.~\cite{tsang19} but
not elaborated there. References~\cite{gefen19,mouradian21} point out
the correspondence between incoherent imaging and noise spectroscopy
more explicitly, although they assume a low dimension for the Hilbert
space and somewhat different parametric models. A more recent
outstanding work by G\'orecki, Riccardi, and Maccone \cite{gorecki22}
also notices the correspondence and also uses the convexity of the
Helstrom information to derive a quantum bound for a random
displacement model with one optical mode. They discovered
independently that unsqueezing before photon counting is optimal for a
squeezed input state and superior to homodyne detection. Another
outstanding relevant work is Ref.~\cite{shi22} by Shi and Zhuang, who
also discovered independently the optimality and superiority of
unsqueezing and photon counting for a somewhat different random
displacement model, which can be obtained by applying a rotating-wave
approximation to the unitary given by Eq.~(\ref{UX}) and imposing a
thermal channel.
References~\cite{gefen19,mouradian21,gorecki22,shi22} all do not
consider the detection problem and are not aware of the prior
Refs.~\cite{tsang_nair,ng16}.

As there exist many other results in quantum-inspired superresolution
that have not yet been translated to noise spectroscopy, and vice
versa, the correspondence between the two models should have a lot
more to give. 

\section*{Acknowledgment}
This research is supported by the National Research Foundation (NRF)
Singapore, under its Quantum Engineering Programme (Grant No.~QEP-P7).

\appendix

\section{\label{app_est}Quantum estimation theory}

Let $\{\rho_\theta:\theta \in \Theta\}$ be a family of density
operators.  Given a parameter $\theta$ and after a measurement modeled
by a positive operator-valued measure (POVM) $\mathcal E$, the
probability measure $Q_\theta$ for the measurement outcome $\lambda$
being in a set $A$ is given by \cite{holevo11}
\begin{align}
Q_\theta(A) &= \trace \mathcal E(A) \rho_\theta,
\end{align}
where $\trace$ is the operator trace. Let the parameter of interest be
a real scalar $\beta(\theta)$ and an estimator be
$\check\beta(\lambda)$.  The mean-square error is defined as
\begin{align}
\MSE(\theta) &\equiv 
\expect_\theta\BK{\Bk{\check\beta-\beta(\theta)}^2}
\\
&= \int dQ_\theta(\lambda) \Bk{\check\beta(\lambda)-\beta(\theta)}^2,
\end{align}
where $\expect_\theta$ denotes the expectation given $\theta$. Let
$\theta$ be a real scalar and suppose that each $Q_\theta$ possesses a
probability density $f_\theta(\lambda)$ with respect to a
$\theta$-independent reference measure $\nu$. Assuming the local
unbiased condition for the estimator given by
\begin{align}
\expect_\theta\bk{\check\beta} &= \beta(\theta),
\\
\int d\nu(\lambda) \check\beta(\lambda) \partial f_\theta(\lambda)
&= \expect_\theta\bk{\check\beta \partial\ln f_\theta} = \partial \beta(\theta),
\\
\partial &\equiv \pdv{}{\theta},
\end{align}
the classical Cram\'er-Rao bound is
\begin{align}
\MSE(\theta) &\ge \frac{[\partial\beta(\theta)]^2}{J(\theta)},
&
J(\theta) &\equiv 
\expect_\theta\Bk{\bk{\partial \ln f_\theta}^2},
\end{align}
where $J(\theta)$ is called the Fisher information.  The bound is also
achievable with the maximum-likelihood estimator in an asymptotic
sense and can be generalized for more relaxed conditions
\cite{vantrees}.

A quantum bound on $J$ for any POVM $\mathcal E$ is \cite{hayashi}
\begin{align}
J(\theta) &\le K(\theta) \equiv \trace \rho_\theta L_\theta^2,
\end{align}
where $K(\theta)$ is the Helstrom information \cite{helstrom},
$L_\theta$ is a solution to
\begin{align}
\partial \rho_\theta &= \rho_\theta \circ L_\theta,
\end{align}
and $\circ$ denotes the Jordan product of two operators defined as
\begin{align}
A \circ B &\equiv \frac{1}{2}\bk{AB+BA}.
\end{align}
Generalization of these results for a multidimensional parameter can
be done by considering the matrix versions of the information quantities
\cite{hayashi} or adopting the parametric-submodel approach
\cite{tsang20,gross20,tsang21a}.

\section{\label{app_qisr}Quantum-inspired superresolution}
Consider the incoherent imaging model given by Eqs.~(\ref{rho}) and
(\ref{UX_imaging}) in one dimension ($M = 1$). For the estimation of
the separation between two point sources, the probability measure
$P_\theta$ as a function of the separation $\theta \in \mathbb R$ can
be modeled as
\begin{align}
P_\theta &= \frac{1}{2}\bk{\delta_{-\theta/2} + \delta_{\theta/2}},
\end{align}
where
\begin{align}
\delta_x(A) &\equiv \begin{cases}1, & x \in A,\\
0, & \textrm{otherwise}
\end{cases}
\end{align}
is the Dirac measure for a unit point mass at position $x$.  With
direct imaging, which can be modeled as a measurement of the
continuous photon position, the Fisher information $J(\theta)$ is
roughly constant for large $\theta$ relative to Rayleigh's criterion,
but it decreases when $\theta$ becomes sub-Rayleigh and drops to zero
when $\theta =0$.  To distinguish this soft penalty to the Fisher
information from the more heuristic Rayleigh's criterion,
Ref.~\cite{tnl} calls the penalty \emph{Rayleigh's curse}.

Unlike the direct-imaging Fisher information under Rayleigh's curse,
the Helstrom information $K(\theta)$ for this problem is constant
regardless of $\theta$, meaning that the ultimate information in the
photons is substantially higher than the direct-imaging information
for sub-Rayleigh separations. Moreover, a measurement in the discrete
Hermite-Gaussian basis called SPADE has a Fisher information that
coincides with the Helstrom information for all $\theta$, meaning that
SPADE is an optimal measurement and can be far superior to direct
imaging for sub-Rayleigh separations \cite{tnl}.

A similar scenario plays out when one attempts to estimate the size
of an object, for which the probability density of $P_\theta$
can be expressed as
\begin{align}
\dv{P_\theta}{X} &= \frac{1}{\theta} w\qty(\frac{X}{\theta})
\label{W}
\end{align}
in terms of a known function $w$ \cite{tsang17,dutton19}. For this
problem, Rayleigh's curse can also be observed in the direct-imaging
information, while the Helstrom information and the SPADE information
approach a nonzero constant for sub-Rayleigh separations.

References~\cite{tsang17,tsang18a,tsang19,zhou19,tsang19b,tsang21a,tsang22}
study the moment estimation problem, assuming that $\theta = P$,
$\Theta$ is the set of all probability meaures, while the parameter of
interest is
\begin{align}
\beta(P) &= \int dP(X) b(X),
\end{align}
such as the second moment with $b(X) = X^2$. As the parameter space is
now infinite-dimensional, the theory becomes much more formidable.
Yet, with some effort, it can still be shown that SPADE enjoys a
significant superiority over direct imaging for sub-Rayleigh object
sizes and is close to quantum-optimal.

\section{\label{app_ec}Extended convexity}
As the Helstrom information $K(\theta)$ is often difficult to compute,
especially if the density operators are mixed, one may have to settle
for looser bounds. If the density operator can be expressed as the
mixture
\begin{align}
\rho_\theta &= \int dP_\theta(Z)  \sigma_\theta(Z),
\end{align}
where $\sigma_\theta(Z)$ is a density operator conditioned on the
random element $Z$, then a useful bound on $K(\theta)$ called the
extended convexity is \cite{alipour,ng16}
\begin{align}
K(\theta) \le \tilde K(\theta) &\equiv 
\int dP_\theta(Z) K[\sigma_\theta(Z)] + J[P_\theta],
\label{ec_gen}
\end{align}
where $K[\sigma_\theta(Z)]$ is the Helstrom information in terms of
$\sigma_\theta(Z)$ and $J[P_\theta]$ is the Fisher information in
terms of $P_\theta$.

To offer some intuition about the use of the extended-convexity bound
in Ref.~\cite{ng16}, consider the model given by Eqs.~(\ref{rho}) and
(\ref{UX_imaging}) in one dimension ($M = 1$) for simplicity.  Let
\begin{align}
dP_\theta &= \frac{1}{\sqrt{2\pi v_X(\theta)}}
\exp[-\frac{X^2}{2 v_X(\theta)}] dX,
\end{align}
such that the variance $v_X(\theta)$ of the displacement $X$ depends
on the unknown parameter. A trick to derive a good bound is to change
the random variable to $Z = X/\gamma(\theta)$ in terms of a
judiciously chosen function $\gamma(\theta)$, such that
\begin{align}
dP_\theta &= \frac{\gamma(\theta)}{\sqrt{2\pi v_X(\theta)}}
\exp[-\frac{\gamma(\theta)^2 Z^2}{2 v_X(\theta)}] dZ,
\\
\sigma_\theta(Z) &= U_{\gamma(\theta) Z}\ket{\psi}\bra{\psi} 
U_{\gamma(\theta) Z}^\dagger.
\end{align}
These expressions lead to
\begin{align}
K[\sigma_\theta(Z)] &= 4v_k (\partial \gamma)^2 Z^2,
\\
v_k &\equiv \bra{\psi}k^2\ket{\psi} - (\bra{\psi}k\ket{\psi})^2,
\\
\int dP_\theta(Z)K[\sigma_\theta(Z)]
&= 4 v_k v_X (\partial \ln \gamma)^2,
\label{K_sigma}
\\
J[P_\theta] &= \frac{1}{2}\Bk{\partial \ln\bk{\frac{v_X}{\gamma^2}}}^2,
\\
\tilde K(\theta) &= 
4 v_k v_X (\partial \ln \gamma)^2 \nonumber\\
&\quad +  \frac{1}{2}\bk{\partial \ln v_X - 2 \partial \ln \gamma}^2,
\end{align}
where Eq.~(\ref{K_sigma}) assumes that $v_k$ and $\gamma$ do not
depend on the random variable. Picking
\begin{align}
\partial \ln \gamma = \frac{\partial \ln v_X}{2 + 4 v_k v_X}
\end{align}
hence leads to
\begin{align}
\tilde K(\theta) &= \frac{(\partial \ln v_X)^2 }{2 + 1/(v_kv_X)},
\end{align}
which resembles Eq.~(\ref{ec}). The derivation of Eq.~(\ref{ec}) in
Ref.~\cite{ng16} indeed follows a similar procedure.

If $P_\theta$ is not Gaussian, a bound may still be obtained by using
the convexity of the Helstrom information; Section~6 of
Ref.~\cite{tsang17} uses the convexity to derive a quantum limit to
object-size estimation in the context of imaging and shows that SPADE
can approach the limit. The trick is to change the variable in
Eq.~(\ref{W}) to $Z = X/\theta$, leading to
\begin{align}
\rho_\theta &= \int dZ w(Z) \sigma_\theta(Z),
&
\sigma_\theta(Z) &= U_{\theta Z}\ket{\psi}\bra{\psi} U_{\theta Z}^\dagger.
\end{align}
As $w(Z)$ no longer depends on $\theta$, Eq.~(\ref{ec_gen}) gives the
convexity bound
\begin{align}
\tilde K(\theta) = \int dZ w(Z) K[\sigma_\theta(Z)]
= 4 v_k \int dZ w(Z) Z^2,
\end{align}
which is independent of $\theta$. It turns out that SPADE
can achieve this bound in the limit of $\theta \to 0$ \cite{tsang17}.
Reference~\cite{gorecki22} has obtained similar results.

\section{\label{app_det}Quantum detection theory}
Assume two hypotheses $\Theta = \{0,1\}$. Let $A$ be the set of
measurement outcomes with which one decides on $\theta = 0$ and $A^c$
be the set with which one decides on $\theta = 1$. Then the type-I and
type-II error probabilities are, respectively,
\begin{align}
Q_0(A^c) &= \trace \mathcal E(A^c)\rho_0,
&
Q_1(A) &= \trace \mathcal E(A)\rho_1.
\end{align}
Let $\pi_\theta$ be the prior probability of the hypothesis $\theta$.
Then the average error probability is
\begin{align}
P_e &\equiv \pi_0Q_0(A^c)  + \pi_1 Q_1(A).
\end{align}
Given a measurement, the classical detection problem is to choose a
set $A$ that minimizes the error probabilities. In particular, $P_e$
can be minimized by a likelihood-ratio test \cite{vantrees}.  Denote
this minimum $P_e$ by $P_{e,\textrm{min}}$ and assume
$\pi_0 = \pi_1 = 1/2$ for simplicity. A useful lower bound on
$P_{e,\textrm{min}}$ is \cite{kailath67}
\begin{align}
P_e &\ge P_{e,\textrm{min}} \ge \frac{1}{2}\bk{1-\sqrt{1-B^2}},
\end{align}
where 
\begin{align}
B &\equiv \int d\nu(\lambda) \sqrt{f_0(\lambda) f_1(\lambda)}
\end{align}
is the Bhattacharyya coefficient.  Another useful bound is the
Chernoff bound given by \cite{vantrees}
\begin{align}
P_{e,\textrm{min}} &\le \frac{1}{2}\exp(-\xi),
\\
\xi &\equiv -\ln \inf_{0\le s \le 1}
\int d\nu(\lambda) [f_0(\lambda)]^{1-s} [f_1(\lambda)]^s.
\end{align}
The error exponent $-\ln P_{e,\textrm{min}}$ approaches the Chernoff
exponent $\xi$ in an asymptotic sense \cite{levy}.

For any POVM $\mathcal E$, a quantum lower bound on $B$ is
\cite{hayashi}
\begin{align}
B &\ge F \equiv \trace \sqrt{\sqrt{\rho_0}\rho_1\sqrt{\rho_0}},
\end{align}
where $F$ is the Uhlmann fidelity, while a quantum upper bound on
$\xi$ is \cite{audenaert08}
\begin{align}
\xi &\le \zeta \equiv 
-\ln \inf_{0\le s \le 1}\trace\qty(\rho_0^{1-s}\rho_1^s).
\end{align}

\bibliography{research2}

\begin{thebibliography}{48}%
\makeatletter
\providecommand \@ifxundefined [1]{%
 \@ifx{#1\undefined}
}%
\providecommand \@ifnum [1]{%
 \ifnum #1\expandafter \@firstoftwo
 \else \expandafter \@secondoftwo
 \fi
}%
\providecommand \@ifx [1]{%
 \ifx #1\expandafter \@firstoftwo
 \else \expandafter \@secondoftwo
 \fi
}%
\providecommand \natexlab [1]{#1}%
\providecommand \enquote  [1]{``#1''}%
\providecommand \bibnamefont  [1]{#1}%
\providecommand \bibfnamefont [1]{#1}%
\providecommand \citenamefont [1]{#1}%
\providecommand \href@noop [0]{\@secondoftwo}%
\providecommand \href [0]{\begingroup \@sanitize@url \@href}%
\providecommand \@href[1]{\@@startlink{#1}\@@href}%
\providecommand \@@href[1]{\endgroup#1\@@endlink}%
\providecommand \@sanitize@url [0]{\catcode `\\12\catcode `\$12\catcode
  `\&12\catcode `\#12\catcode `\^12\catcode `\_12\catcode `\%12\relax}%
\providecommand \@@startlink[1]{}%
\providecommand \@@endlink[0]{}%
\providecommand \url  [0]{\begingroup\@sanitize@url \@url }%
\providecommand \@url [1]{\endgroup\@href {#1}{\urlprefix }}%
\providecommand \urlprefix  [0]{URL }%
\providecommand \Eprint [0]{\href }%
\providecommand \doibase [0]{http://dx.doi.org/}%
\providecommand \selectlanguage [0]{\@gobble}%
\providecommand \bibinfo  [0]{\@secondoftwo}%
\providecommand \bibfield  [0]{\@secondoftwo}%
\providecommand \translation [1]{[#1]}%
\providecommand \BibitemOpen [0]{}%
\providecommand \bibitemStop [0]{}%
\providecommand \bibitemNoStop [0]{.\EOS\space}%
\providecommand \EOS [0]{\spacefactor3000\relax}%
\providecommand \BibitemShut  [1]{\csname bibitem#1\endcsname}%
\let\auto@bib@innerbib\@empty
\bibitem [{\citenamefont {Tsang}\ \emph {et~al.}(2016)\citenamefont {Tsang},
  \citenamefont {Nair},\ and\ \citenamefont {Lu}}]{tnl}%
  \BibitemOpen
  \bibfield  {author} {\bibinfo {author} {\bibfnamefont {Mankei}\ \bibnamefont
  {Tsang}}, \bibinfo {author} {\bibfnamefont {Ranjith}\ \bibnamefont {Nair}}, \
  and\ \bibinfo {author} {\bibfnamefont {Xiao-Ming}\ \bibnamefont {Lu}},\
  }\bibfield  {title} {\enquote {\bibinfo {title} {Quantum theory of
  superresolution for two incoherent optical point sources},}\ }\href {\doibase
  10.1103/PhysRevX.6.031033} {\bibfield  {journal} {\bibinfo  {journal}
  {Physical Review X}\ }\textbf {\bibinfo {volume} {6}},\ \bibinfo {pages}
  {031033} (\bibinfo {year} {2016})}\BibitemShut {NoStop}%
\bibitem [{\citenamefont {Tsang}(2019{\natexlab{a}})}]{tsang19a}%
  \BibitemOpen
  \bibfield  {author} {\bibinfo {author} {\bibfnamefont {Mankei}\ \bibnamefont
  {Tsang}},\ }\bibfield  {title} {\enquote {\bibinfo {title} {Resolving
  starlight: a quantum perspective},}\ }\href {\doibase
  10.1080/00107514.2020.1736375} {\bibfield  {journal} {\bibinfo  {journal}
  {Contemporary Physics}\ }\textbf {\bibinfo {volume} {60}},\ \bibinfo {pages}
  {279--298} (\bibinfo {year} {2019}{\natexlab{a}})}\BibitemShut {NoStop}%
\bibitem [{\citenamefont {Tsang}\ and\ \citenamefont
  {Nair}(2012)}]{tsang_nair}%
  \BibitemOpen
  \bibfield  {author} {\bibinfo {author} {\bibfnamefont {Mankei}\ \bibnamefont
  {Tsang}}\ and\ \bibinfo {author} {\bibfnamefont {Ranjith}\ \bibnamefont
  {Nair}},\ }\bibfield  {title} {\enquote {\bibinfo {title} {Fundamental
  quantum limits to waveform detection},}\ }\href {\doibase
  10.1103/PhysRevA.86.042115} {\bibfield  {journal} {\bibinfo  {journal}
  {Physical Review A}\ }\textbf {\bibinfo {volume} {86}},\ \bibinfo {pages}
  {042115} (\bibinfo {year} {2012})}\BibitemShut {NoStop}%
\bibitem [{\citenamefont {Ng}\ \emph {et~al.}(2016)\citenamefont {Ng},
  \citenamefont {Ang}, \citenamefont {Wheatley}, \citenamefont {Yonezawa},
  \citenamefont {Furusawa}, \citenamefont {Huntington},\ and\ \citenamefont
  {Tsang}}]{ng16}%
  \BibitemOpen
  \bibfield  {author} {\bibinfo {author} {\bibfnamefont {Shilin}\ \bibnamefont
  {Ng}}, \bibinfo {author} {\bibfnamefont {Shan~Zheng}\ \bibnamefont {Ang}},
  \bibinfo {author} {\bibfnamefont {Trevor~A.}\ \bibnamefont {Wheatley}},
  \bibinfo {author} {\bibfnamefont {Hidehiro}\ \bibnamefont {Yonezawa}},
  \bibinfo {author} {\bibfnamefont {Akira}\ \bibnamefont {Furusawa}}, \bibinfo
  {author} {\bibfnamefont {Elanor~H.}\ \bibnamefont {Huntington}}, \ and\
  \bibinfo {author} {\bibfnamefont {Mankei}\ \bibnamefont {Tsang}},\ }\bibfield
   {title} {\enquote {\bibinfo {title} {Spectrum analysis with quantum
  dynamical systems},}\ }\href {\doibase 10.1103/PhysRevA.93.042121} {\bibfield
   {journal} {\bibinfo  {journal} {Physical Review A}\ }\textbf {\bibinfo
  {volume} {93}},\ \bibinfo {pages} {042121} (\bibinfo {year}
  {2016})}\BibitemShut {NoStop}%
\bibitem [{\citenamefont {Budker}\ and\ \citenamefont
  {Romalis}(2007)}]{budker07}%
  \BibitemOpen
  \bibfield  {author} {\bibinfo {author} {\bibfnamefont {Dmitry}\ \bibnamefont
  {Budker}}\ and\ \bibinfo {author} {\bibfnamefont {Michael}\ \bibnamefont
  {Romalis}},\ }\bibfield  {title} {\enquote {\bibinfo {title} {Optical
  magnetometry},}\ }\href {\doibase 10.1038/nphys566} {\bibfield  {journal}
  {\bibinfo  {journal} {Nature Physics}\ }\textbf {\bibinfo {volume} {3}},\
  \bibinfo {pages} {227--234} (\bibinfo {year} {2007})}\BibitemShut {NoStop}%
\bibitem [{\citenamefont {Backes}\ \emph {et~al.}(2021)\citenamefont {Backes},
  \citenamefont {Palken}, \citenamefont {Kenany}, \citenamefont {Brubaker},
  \citenamefont {Cahn}, \citenamefont {Droster}, \citenamefont {Hilton},
  \citenamefont {Ghosh}, \citenamefont {Jackson}, \citenamefont {Lamoreaux},
  \citenamefont {Leder}, \citenamefont {Lehnert}, \citenamefont {Lewis},
  \citenamefont {Malnou}, \citenamefont {Maruyama}, \citenamefont {Rapidis},
  \citenamefont {Simanovskaia}, \citenamefont {Singh}, \citenamefont {Speller},
  \citenamefont {Urdinaran}, \citenamefont {Vale}, \citenamefont {van
  Assendelft}, \citenamefont {van Bibber},\ and\ \citenamefont
  {Wang}}]{backes21}%
  \BibitemOpen
  \bibfield  {author} {\bibinfo {author} {\bibfnamefont {K.~M.}\ \bibnamefont
  {Backes}}, \bibinfo {author} {\bibfnamefont {D.~A.}\ \bibnamefont {Palken}},
  \bibinfo {author} {\bibfnamefont {S.~Al}\ \bibnamefont {Kenany}}, \bibinfo
  {author} {\bibfnamefont {B.~M.}\ \bibnamefont {Brubaker}}, \bibinfo {author}
  {\bibfnamefont {S.~B.}\ \bibnamefont {Cahn}}, \bibinfo {author}
  {\bibfnamefont {A.}~\bibnamefont {Droster}}, \bibinfo {author} {\bibfnamefont
  {Gene~C.}\ \bibnamefont {Hilton}}, \bibinfo {author} {\bibfnamefont {Sumita}\
  \bibnamefont {Ghosh}}, \bibinfo {author} {\bibfnamefont {H.}~\bibnamefont
  {Jackson}}, \bibinfo {author} {\bibfnamefont {S.~K.}\ \bibnamefont
  {Lamoreaux}}, \bibinfo {author} {\bibfnamefont {A.~F.}\ \bibnamefont
  {Leder}}, \bibinfo {author} {\bibfnamefont {K.~W.}\ \bibnamefont {Lehnert}},
  \bibinfo {author} {\bibfnamefont {S.~M.}\ \bibnamefont {Lewis}}, \bibinfo
  {author} {\bibfnamefont {M.}~\bibnamefont {Malnou}}, \bibinfo {author}
  {\bibfnamefont {R.~H.}\ \bibnamefont {Maruyama}}, \bibinfo {author}
  {\bibfnamefont {N.~M.}\ \bibnamefont {Rapidis}}, \bibinfo {author}
  {\bibfnamefont {M.}~\bibnamefont {Simanovskaia}}, \bibinfo {author}
  {\bibfnamefont {Sukhman}\ \bibnamefont {Singh}}, \bibinfo {author}
  {\bibfnamefont {D.~H.}\ \bibnamefont {Speller}}, \bibinfo {author}
  {\bibfnamefont {I.}~\bibnamefont {Urdinaran}}, \bibinfo {author}
  {\bibfnamefont {Leila~R.}\ \bibnamefont {Vale}}, \bibinfo {author}
  {\bibfnamefont {E.~C.}\ \bibnamefont {van Assendelft}}, \bibinfo {author}
  {\bibfnamefont {K.}~\bibnamefont {van Bibber}}, \ and\ \bibinfo {author}
  {\bibfnamefont {H.}~\bibnamefont {Wang}},\ }\bibfield  {title} {\enquote
  {\bibinfo {title} {A quantum enhanced search for dark matter axions},}\
  }\href {\doibase 10.1038/s41586-021-03226-7} {\bibfield  {journal} {\bibinfo
  {journal} {Nature}\ }\textbf {\bibinfo {volume} {590}},\ \bibinfo {pages}
  {238--242} (\bibinfo {year} {2021})}\BibitemShut {NoStop}%
\bibitem [{\citenamefont {Holevo}(2011)}]{holevo11}%
  \BibitemOpen
  \bibfield  {author} {\bibinfo {author} {\bibfnamefont {Alexander~S.}\
  \bibnamefont {Holevo}},\ }\href {\doibase 10.1007/978-88-7642-378-9} {\emph
  {\bibinfo {title} {Probabilistic and Statistical Aspects of Quantum
  Theory}}}\ (\bibinfo  {publisher} {Scuola Normale Superiore Pisa},\ \bibinfo
  {address} {Pisa, Italy},\ \bibinfo {year} {2011})\BibitemShut {NoStop}%
\bibitem [{\citenamefont {Helstrom}(1976)}]{helstrom}%
  \BibitemOpen
  \bibfield  {author} {\bibinfo {author} {\bibfnamefont {Carl~W.}\ \bibnamefont
  {Helstrom}},\ }\href
  {http://www.sciencedirect.com/science/bookseries/00765392/123} {\emph
  {\bibinfo {title} {Quantum Detection and Estimation Theory}}}\ (\bibinfo
  {publisher} {Academic Press},\ \bibinfo {address} {New York},\ \bibinfo
  {year} {1976})\BibitemShut {NoStop}%
\bibitem [{\citenamefont {Lu}\ \emph {et~al.}(2018)\citenamefont {Lu},
  \citenamefont {Krovi}, \citenamefont {Nair}, \citenamefont {Guha},\ and\
  \citenamefont {Shapiro}}]{lu18}%
  \BibitemOpen
  \bibfield  {author} {\bibinfo {author} {\bibfnamefont {Xiao-Ming}\
  \bibnamefont {Lu}}, \bibinfo {author} {\bibfnamefont {Hari}\ \bibnamefont
  {Krovi}}, \bibinfo {author} {\bibfnamefont {Ranjith}\ \bibnamefont {Nair}},
  \bibinfo {author} {\bibfnamefont {Saikat}\ \bibnamefont {Guha}}, \ and\
  \bibinfo {author} {\bibfnamefont {Jeffrey~H.}\ \bibnamefont {Shapiro}},\
  }\bibfield  {title} {\enquote {\bibinfo {title} {Quantum-optimal detection of
  one-versus-two incoherent optical sources with arbitrary separation},}\
  }\href {\doibase 10.1038/s41534-018-0114-y} {\bibfield  {journal} {\bibinfo
  {journal} {npj Quantum Information}\ }\textbf {\bibinfo {volume} {4}},\
  \bibinfo {pages} {64} (\bibinfo {year} {2018})}\BibitemShut {NoStop}%
\bibitem [{\citenamefont {Tsang}(2017)}]{tsang17}%
  \BibitemOpen
  \bibfield  {author} {\bibinfo {author} {\bibfnamefont {Mankei}\ \bibnamefont
  {Tsang}},\ }\bibfield  {title} {\enquote {\bibinfo {title} {Subdiffraction
  incoherent optical imaging via spatial-mode demultiplexing},}\ }\href
  {\doibase 10.1088/1367-2630/aa60ee} {\bibfield  {journal} {\bibinfo
  {journal} {New Journal of Physics}\ }\textbf {\bibinfo {volume} {19}},\
  \bibinfo {pages} {023054} (\bibinfo {year} {2017})}\BibitemShut {NoStop}%
\bibitem [{\citenamefont {Dutton}\ \emph {et~al.}(2019)\citenamefont {Dutton},
  \citenamefont {Kerviche}, \citenamefont {Ashok},\ and\ \citenamefont
  {Guha}}]{dutton19}%
  \BibitemOpen
  \bibfield  {author} {\bibinfo {author} {\bibfnamefont {Zachary}\ \bibnamefont
  {Dutton}}, \bibinfo {author} {\bibfnamefont {Ronan}\ \bibnamefont
  {Kerviche}}, \bibinfo {author} {\bibfnamefont {Amit}\ \bibnamefont {Ashok}},
  \ and\ \bibinfo {author} {\bibfnamefont {Saikat}\ \bibnamefont {Guha}},\
  }\bibfield  {title} {\enquote {\bibinfo {title} {Attaining the quantum limit
  of superresolution in imaging an object's length via predetection
  spatial-mode sorting},}\ }\href {\doibase 10.1103/PhysRevA.99.033847}
  {\bibfield  {journal} {\bibinfo  {journal} {Physical Review A}\ }\textbf
  {\bibinfo {volume} {99}},\ \bibinfo {pages} {033847} (\bibinfo {year}
  {2019})}\BibitemShut {NoStop}%
\bibitem [{\citenamefont {Tsang}(2018)}]{tsang18a}%
  \BibitemOpen
  \bibfield  {author} {\bibinfo {author} {\bibfnamefont {Mankei}\ \bibnamefont
  {Tsang}},\ }\bibfield  {title} {\enquote {\bibinfo {title} {Subdiffraction
  incoherent optical imaging via spatial-mode demultiplexing: {Semiclassical}
  treatment},}\ }\href {\doibase 10.1103/PhysRevA.97.023830} {\bibfield
  {journal} {\bibinfo  {journal} {Physical Review A}\ }\textbf {\bibinfo
  {volume} {97}},\ \bibinfo {pages} {023830} (\bibinfo {year}
  {2018})}\BibitemShut {NoStop}%
\bibitem [{\citenamefont {Tsang}(2019{\natexlab{b}})}]{tsang19}%
  \BibitemOpen
  \bibfield  {author} {\bibinfo {author} {\bibfnamefont {Mankei}\ \bibnamefont
  {Tsang}},\ }\bibfield  {title} {\enquote {\bibinfo {title} {Quantum limit to
  subdiffraction incoherent optical imaging},}\ }\href {\doibase
  10.1103/PhysRevA.99.012305} {\bibfield  {journal} {\bibinfo  {journal}
  {Physical Review A}\ }\textbf {\bibinfo {volume} {99}},\ \bibinfo {pages}
  {012305} (\bibinfo {year} {2019}{\natexlab{b}})}\BibitemShut {NoStop}%
\bibitem [{\citenamefont {Zhou}\ and\ \citenamefont {Jiang}(2019)}]{zhou19}%
  \BibitemOpen
  \bibfield  {author} {\bibinfo {author} {\bibfnamefont {Sisi}\ \bibnamefont
  {Zhou}}\ and\ \bibinfo {author} {\bibfnamefont {Liang}\ \bibnamefont
  {Jiang}},\ }\bibfield  {title} {\enquote {\bibinfo {title} {Modern
  description of {Rayleigh}'s criterion},}\ }\href {\doibase
  10.1103/PhysRevA.99.013808} {\bibfield  {journal} {\bibinfo  {journal}
  {Physical Review A}\ }\textbf {\bibinfo {volume} {99}},\ \bibinfo {pages}
  {013808} (\bibinfo {year} {2019})}\BibitemShut {NoStop}%
\bibitem [{\citenamefont {Tsang}(2019{\natexlab{c}})}]{tsang19b}%
  \BibitemOpen
  \bibfield  {author} {\bibinfo {author} {\bibfnamefont {Mankei}\ \bibnamefont
  {Tsang}},\ }\bibfield  {title} {\enquote {\bibinfo {title} {Semiparametric
  estimation for incoherent optical imaging},}\ }\href {\doibase
  10.1103/PhysRevResearch.1.033006} {\bibfield  {journal} {\bibinfo  {journal}
  {Physical Review Research}\ }\textbf {\bibinfo {volume} {1}},\ \bibinfo
  {pages} {033006} (\bibinfo {year} {2019}{\natexlab{c}})}\BibitemShut
  {NoStop}%
\bibitem [{\citenamefont {Tsang}(2021)}]{tsang21a}%
  \BibitemOpen
  \bibfield  {author} {\bibinfo {author} {\bibfnamefont {Mankei}\ \bibnamefont
  {Tsang}},\ }\bibfield  {title} {\enquote {\bibinfo {title} {Quantum limit to
  subdiffraction incoherent optical imaging. {II}. {A} parametric-submodel
  approach},}\ }\href {\doibase 10.1103/PhysRevA.104.052411} {\bibfield
  {journal} {\bibinfo  {journal} {Physical Review A}\ }\textbf {\bibinfo
  {volume} {104}},\ \bibinfo {pages} {052411} (\bibinfo {year}
  {2021})}\BibitemShut {NoStop}%
\bibitem [{\citenamefont {Tsang}(2022)}]{tsang22}%
  \BibitemOpen
  \bibfield  {author} {\bibinfo {author} {\bibfnamefont {Mankei}\ \bibnamefont
  {Tsang}},\ }\bibfield  {title} {\enquote {\bibinfo {title} {Efficient
  superoscillation measurement for incoherent optical imaging},}\ }\href
  {http://arxiv.org/abs/2010.11084} {\bibfield  {journal} {\bibinfo  {journal}
  {arXiv:2010.11084}\ } (\bibinfo {year} {2022})}\BibitemShut {NoStop}%
\bibitem [{\citenamefont {Tsang}(2013)}]{tsang_open}%
  \BibitemOpen
  \bibfield  {author} {\bibinfo {author} {\bibfnamefont {Mankei}\ \bibnamefont
  {Tsang}},\ }\bibfield  {title} {\enquote {\bibinfo {title} {Quantum metrology
  with open dynamical systems},}\ }\href {\doibase
  10.1088/1367-2630/15/7/073005} {\bibfield  {journal} {\bibinfo  {journal}
  {New Journal of Physics}\ }\textbf {\bibinfo {volume} {15}},\ \bibinfo
  {pages} {073005} (\bibinfo {year} {2013})}\BibitemShut {NoStop}%
\bibitem [{\citenamefont {Tsang}\ \emph {et~al.}(2011)\citenamefont {Tsang},
  \citenamefont {Wiseman},\ and\ \citenamefont {Caves}}]{twc}%
  \BibitemOpen
  \bibfield  {author} {\bibinfo {author} {\bibfnamefont {Mankei}\ \bibnamefont
  {Tsang}}, \bibinfo {author} {\bibfnamefont {Howard~M.}\ \bibnamefont
  {Wiseman}}, \ and\ \bibinfo {author} {\bibfnamefont {Carlton~M.}\
  \bibnamefont {Caves}},\ }\bibfield  {title} {\enquote {\bibinfo {title}
  {Fundamental {Quantum} {Limit} to {Waveform} {Estimation}},}\ }\href
  {\doibase 10.1103/PhysRevLett.106.090401} {\bibfield  {journal} {\bibinfo
  {journal} {Physical Review Letters}\ }\textbf {\bibinfo {volume} {106}},\
  \bibinfo {pages} {090401} (\bibinfo {year} {2011})}\BibitemShut {NoStop}%
\bibitem [{\citenamefont {Nielsen}\ and\ \citenamefont
  {Chuang}(2011)}]{nielsen}%
  \BibitemOpen
  \bibfield  {author} {\bibinfo {author} {\bibfnamefont {Michael~A.}\
  \bibnamefont {Nielsen}}\ and\ \bibinfo {author} {\bibfnamefont {Isaac~L.}\
  \bibnamefont {Chuang}},\ }\href@noop {} {\emph {\bibinfo {title} {Quantum
  Computation and Quantum Information}}}\ (\bibinfo  {publisher} {Cambridge
  University Press},\ \bibinfo {address} {Cambridge},\ \bibinfo {year}
  {2011})\BibitemShut {NoStop}%
\bibitem [{\citenamefont {Aspelmeyer}\ \emph {et~al.}(2014)\citenamefont
  {Aspelmeyer}, \citenamefont {Kippenberg},\ and\ \citenamefont
  {Marquardt}}]{aspelmeyer14}%
  \BibitemOpen
  \bibfield  {author} {\bibinfo {author} {\bibfnamefont {Markus}\ \bibnamefont
  {Aspelmeyer}}, \bibinfo {author} {\bibfnamefont {Tobias~J.}\ \bibnamefont
  {Kippenberg}}, \ and\ \bibinfo {author} {\bibfnamefont {Florian}\
  \bibnamefont {Marquardt}},\ }\bibfield  {title} {\enquote {\bibinfo {title}
  {Cavity optomechanics},}\ }\href {\doibase 10.1103/RevModPhys.86.1391}
  {\bibfield  {journal} {\bibinfo  {journal} {Rev. Mod. Phys.}\ }\textbf
  {\bibinfo {volume} {86}},\ \bibinfo {pages} {1391--1452} (\bibinfo {year}
  {2014})}\BibitemShut {NoStop}%
\bibitem [{\citenamefont {Nimmrichter}\ \emph {et~al.}(2014)\citenamefont
  {Nimmrichter}, \citenamefont {Hornberger},\ and\ \citenamefont
  {Hammerer}}]{nimmrichter}%
  \BibitemOpen
  \bibfield  {author} {\bibinfo {author} {\bibfnamefont {Stefan}\ \bibnamefont
  {Nimmrichter}}, \bibinfo {author} {\bibfnamefont {Klaus}\ \bibnamefont
  {Hornberger}}, \ and\ \bibinfo {author} {\bibfnamefont {Klemens}\
  \bibnamefont {Hammerer}},\ }\bibfield  {title} {\enquote {\bibinfo {title}
  {Optomechanical sensing of spontaneous wave-function collapse},}\ }\href
  {\doibase 10.1103/PhysRevLett.113.020405} {\bibfield  {journal} {\bibinfo
  {journal} {Physical Review Letters}\ }\textbf {\bibinfo {volume} {113}},\
  \bibinfo {pages} {020405} (\bibinfo {year} {2014})}\BibitemShut {NoStop}%
\bibitem [{\citenamefont {Christensen}(2019)}]{christensen19}%
  \BibitemOpen
  \bibfield  {author} {\bibinfo {author} {\bibfnamefont {Nelson}\ \bibnamefont
  {Christensen}},\ }\bibfield  {title} {\enquote {\bibinfo {title} {Stochastic
  gravitational wave backgrounds},}\ }\href {\doibase 10.1088/1361-6633/aae6b5}
  {\bibfield  {journal} {\bibinfo  {journal} {Reports on Progress in Physics}\
  }\textbf {\bibinfo {volume} {82}},\ \bibinfo {pages} {016903} (\bibinfo
  {year} {2019})}\BibitemShut {NoStop}%
\bibitem [{\citenamefont {Gardiner}\ and\ \citenamefont
  {Zoller}(2000)}]{gardiner_zoller}%
  \BibitemOpen
  \bibfield  {author} {\bibinfo {author} {\bibfnamefont {Crispin~W.}\
  \bibnamefont {Gardiner}}\ and\ \bibinfo {author} {\bibfnamefont {Peter}\
  \bibnamefont {Zoller}},\ }\href@noop {} {\emph {\bibinfo {title} {Quantum
  Noise}}},\ \bibinfo {edition} {2nd}\ ed.\ (\bibinfo  {publisher}
  {Springer-Verlag},\ \bibinfo {address} {Berlin},\ \bibinfo {year}
  {2000})\BibitemShut {NoStop}%
\bibitem [{\citenamefont {Shapiro}(2009)}]{shapiro09}%
  \BibitemOpen
  \bibfield  {author} {\bibinfo {author} {\bibfnamefont {Jeffrey~H.}\
  \bibnamefont {Shapiro}},\ }\bibfield  {title} {\enquote {\bibinfo {title}
  {The quantum theory of optical communications},}\ }\href {\doibase
  10.1109/JSTQE.2009.2024959} {\bibfield  {journal} {\bibinfo  {journal} {IEEE
  Journal of Selected Topics in Quantum Electronics}\ }\textbf {\bibinfo
  {volume} {15}},\ \bibinfo {pages} {1547--1569} (\bibinfo {year}
  {2009})}\BibitemShut {NoStop}%
\bibitem [{\citenamefont {{Van Trees}}(2001)}]{vantrees}%
  \BibitemOpen
  \bibfield  {author} {\bibinfo {author} {\bibfnamefont {Harry~L.}\
  \bibnamefont {{Van Trees}}},\ }\href@noop {} {\emph {\bibinfo {title}
  {Detection, Estimation, and Modulation Theory, Part I.}}}\ (\bibinfo
  {publisher} {John Wiley \& Sons},\ \bibinfo {address} {New York},\ \bibinfo
  {year} {2001})\BibitemShut {NoStop}%
\bibitem [{\citenamefont {Shumway}\ and\ \citenamefont
  {Stoffer}(2017)}]{shumway_stoffer}%
  \BibitemOpen
  \bibfield  {author} {\bibinfo {author} {\bibfnamefont {Robert~H.}\
  \bibnamefont {Shumway}}\ and\ \bibinfo {author} {\bibfnamefont {David~S.}\
  \bibnamefont {Stoffer}},\ }\href {\doibase 10.1007/978-3-319-52452-8} {\emph
  {\bibinfo {title} {Time Series Analysis and Its Applications}}}\ (\bibinfo
  {publisher} {Springer},\ \bibinfo {address} {Cham, Switzerland},\ \bibinfo
  {year} {2017})\BibitemShut {NoStop}%
\bibitem [{\citenamefont {Hayashi}(2017)}]{hayashi}%
  \BibitemOpen
  \bibfield  {author} {\bibinfo {author} {\bibfnamefont {Masahito}\
  \bibnamefont {Hayashi}},\ }\href {\doibase 10.1007/978-3-662-49725-8} {\emph
  {\bibinfo {title} {Quantum {I}nformation {T}heory: {M}athematical
  {F}oundation}}},\ \bibinfo {edition} {2nd}\ ed.\ (\bibinfo  {publisher}
  {Springer},\ \bibinfo {address} {Berlin},\ \bibinfo {year}
  {2017})\BibitemShut {NoStop}%
\bibitem [{\citenamefont {Alipour}\ and\ \citenamefont
  {Rezakhani}(2015)}]{alipour}%
  \BibitemOpen
  \bibfield  {author} {\bibinfo {author} {\bibfnamefont {S.}~\bibnamefont
  {Alipour}}\ and\ \bibinfo {author} {\bibfnamefont {A.~T.}\ \bibnamefont
  {Rezakhani}},\ }\bibfield  {title} {\enquote {\bibinfo {title} {Extended
  convexity of quantum {F}isher information in quantum metrology},}\ }\href
  {\doibase 10.1103/PhysRevA.91.042104} {\bibfield  {journal} {\bibinfo
  {journal} {Physical Review A}\ }\textbf {\bibinfo {volume} {91}},\ \bibinfo
  {pages} {042104} (\bibinfo {year} {2015})}\BibitemShut {NoStop}%
\bibitem [{\citenamefont {{\v R}eh{\'a}{\v c}ek}\ \emph
  {et~al.}(2017)\citenamefont {{\v R}eh{\'a}{\v c}ek}, \citenamefont
  {Pa{\'u}r}, \citenamefont {Stoklasa}, \citenamefont {Hradil},\ and\
  \citenamefont {S{\'a}nchez-Soto}}]{rehacek17}%
  \BibitemOpen
  \bibfield  {author} {\bibinfo {author} {\bibfnamefont {J.}~\bibnamefont {{\v
  R}eh{\'a}{\v c}ek}}, \bibinfo {author} {\bibfnamefont {M.}~\bibnamefont
  {Pa{\'u}r}}, \bibinfo {author} {\bibfnamefont {B.}~\bibnamefont {Stoklasa}},
  \bibinfo {author} {\bibfnamefont {Z.}~\bibnamefont {Hradil}}, \ and\ \bibinfo
  {author} {\bibfnamefont {L.~L.}\ \bibnamefont {S{\'a}nchez-Soto}},\
  }\bibfield  {title} {\enquote {\bibinfo {title} {Optimal measurements for
  resolution beyond the {Rayleigh} limit},}\ }\href {\doibase
  10.1364/OL.42.000231} {\bibfield  {journal} {\bibinfo  {journal} {Optics
  Letters}\ }\textbf {\bibinfo {volume} {42}},\ \bibinfo {pages} {231--234}
  (\bibinfo {year} {2017})}\BibitemShut {NoStop}%
\bibitem [{\citenamefont {Caves}(1981)}]{caves81}%
  \BibitemOpen
  \bibfield  {author} {\bibinfo {author} {\bibfnamefont {Carlton~M.}\
  \bibnamefont {Caves}},\ }\bibfield  {title} {\enquote {\bibinfo {title}
  {Quantum-mechanical noise in an interferometer},}\ }\href {\doibase
  10.1103/PhysRevD.23.1693} {\bibfield  {journal} {\bibinfo  {journal}
  {Physical Review D}\ }\textbf {\bibinfo {volume} {23}},\ \bibinfo {pages}
  {1693--1708} (\bibinfo {year} {1981})}\BibitemShut {NoStop}%
\bibitem [{\citenamefont {Kimble}\ \emph {et~al.}(2001)\citenamefont {Kimble},
  \citenamefont {Levin}, \citenamefont {Matsko}, \citenamefont {Thorne},\ and\
  \citenamefont {Vyatchanin}}]{klmtv}%
  \BibitemOpen
  \bibfield  {author} {\bibinfo {author} {\bibfnamefont {H.~J.}\ \bibnamefont
  {Kimble}}, \bibinfo {author} {\bibfnamefont {Yuri}\ \bibnamefont {Levin}},
  \bibinfo {author} {\bibfnamefont {Andrey~B.}\ \bibnamefont {Matsko}},
  \bibinfo {author} {\bibfnamefont {Kip~S.}\ \bibnamefont {Thorne}}, \ and\
  \bibinfo {author} {\bibfnamefont {Sergey~P.}\ \bibnamefont {Vyatchanin}},\
  }\bibfield  {title} {\enquote {\bibinfo {title} {Conversion of conventional
  gravitational-wave interferometers into quantum nondemolition interferometers
  by modifying their input and/or output optics},}\ }\href {\doibase
  10.1103/PhysRevD.65.022002} {\bibfield  {journal} {\bibinfo  {journal}
  {Physical Review D}\ }\textbf {\bibinfo {volume} {65}},\ \bibinfo {pages}
  {022002} (\bibinfo {year} {2001})}\BibitemShut {NoStop}%
\bibitem [{\citenamefont {Tsang}\ and\ \citenamefont {Caves}(2010)}]{qnc}%
  \BibitemOpen
  \bibfield  {author} {\bibinfo {author} {\bibfnamefont {Mankei}\ \bibnamefont
  {Tsang}}\ and\ \bibinfo {author} {\bibfnamefont {Carlton~M.}\ \bibnamefont
  {Caves}},\ }\bibfield  {title} {\enquote {\bibinfo {title} {Coherent
  {Quantum}-{Noise} {Cancellation} for {Optomechanical} {Sensors}},}\ }\href
  {\doibase 10.1103/PhysRevLett.105.123601} {\bibfield  {journal} {\bibinfo
  {journal} {Physical Review Letters}\ }\textbf {\bibinfo {volume} {105}},\
  \bibinfo {pages} {123601} (\bibinfo {year} {2010})}\BibitemShut {NoStop}%
\bibitem [{\citenamefont {Shapiro}(1998)}]{shapiro98}%
  \BibitemOpen
  \bibfield  {author} {\bibinfo {author} {\bibfnamefont {Jeffrey~H.}\
  \bibnamefont {Shapiro}},\ }\bibfield  {title} {\enquote {\bibinfo {title}
  {Quantum measurement eigenkets for continuous-time direct detection},}\
  }\href {http://stacks.iop.org/1355-5111/10/i=3/a=014} {\bibfield  {journal}
  {\bibinfo  {journal} {Quantum and Semiclassical Optics: Journal of the
  European Optical Society Part B}\ }\textbf {\bibinfo {volume} {10}},\
  \bibinfo {pages} {567} (\bibinfo {year} {1998})}\BibitemShut {NoStop}%
\bibitem [{\citenamefont {Whittle}(1953)}]{whittle}%
  \BibitemOpen
  \bibfield  {author} {\bibinfo {author} {\bibfnamefont {P.}~\bibnamefont
  {Whittle}},\ }\bibfield  {title} {\enquote {\bibinfo {title} {The analysis of
  multiple stationary time series},}\ }\href
  {http://www.jstor.org/stable/2983728} {\bibfield  {journal} {\bibinfo
  {journal} {Journal of the Royal Statistical Society. Series B
  (Methodological)}\ }\textbf {\bibinfo {volume} {15}},\ \bibinfo {pages} {pp.
  125--139} (\bibinfo {year} {1953})}\BibitemShut {NoStop}%
\bibitem [{\citenamefont {Audenaert}\ \emph {et~al.}(2008)\citenamefont
  {Audenaert}, \citenamefont {Nussbaum}, \citenamefont {Szko{\l}a},\ and\
  \citenamefont {Verstraete}}]{audenaert08}%
  \BibitemOpen
  \bibfield  {author} {\bibinfo {author} {\bibfnamefont {K.M.R.}\ \bibnamefont
  {Audenaert}}, \bibinfo {author} {\bibfnamefont {M.}~\bibnamefont {Nussbaum}},
  \bibinfo {author} {\bibfnamefont {A.}~\bibnamefont {Szko{\l}a}}, \ and\
  \bibinfo {author} {\bibfnamefont {F.}~\bibnamefont {Verstraete}},\ }\bibfield
   {title} {\enquote {\bibinfo {title} {Asymptotic error rates in quantum
  hypothesis testing},}\ }\href {\doibase 10.1007/s00220-008-0417-5} {\bibfield
   {journal} {\bibinfo  {journal} {Communications in Mathematical Physics}\
  }\textbf {\bibinfo {volume} {279}},\ \bibinfo {pages} {251--283} (\bibinfo
  {year} {2008})}\BibitemShut {NoStop}%
\bibitem [{\citenamefont {Huang}\ and\ \citenamefont {Lupo}(2021)}]{huang21}%
  \BibitemOpen
  \bibfield  {author} {\bibinfo {author} {\bibfnamefont {Zixin}\ \bibnamefont
  {Huang}}\ and\ \bibinfo {author} {\bibfnamefont {Cosmo}\ \bibnamefont
  {Lupo}},\ }\bibfield  {title} {\enquote {\bibinfo {title} {Quantum hypothesis
  testing for exoplanet detection},}\ }\href {\doibase
  10.1103/PhysRevLett.127.130502} {\bibfield  {journal} {\bibinfo  {journal}
  {Physical Review Letters}\ }\textbf {\bibinfo {volume} {127}},\ \bibinfo
  {pages} {130502} (\bibinfo {year} {2021})}\BibitemShut {NoStop}%
\bibitem [{\citenamefont {Zanforlin}\ \emph {et~al.}(2022)\citenamefont
  {Zanforlin}, \citenamefont {Lupo}, \citenamefont {Connolly}, \citenamefont
  {Kok}, \citenamefont {Buller},\ and\ \citenamefont {Huang}}]{zanforlin22}%
  \BibitemOpen
  \bibfield  {author} {\bibinfo {author} {\bibfnamefont {Ugo}\ \bibnamefont
  {Zanforlin}}, \bibinfo {author} {\bibfnamefont {Cosmo}\ \bibnamefont {Lupo}},
  \bibinfo {author} {\bibfnamefont {Peter W.~R.}\ \bibnamefont {Connolly}},
  \bibinfo {author} {\bibfnamefont {Pieter}\ \bibnamefont {Kok}}, \bibinfo
  {author} {\bibfnamefont {Gerald~S.}\ \bibnamefont {Buller}}, \ and\ \bibinfo
  {author} {\bibfnamefont {Zixin}\ \bibnamefont {Huang}},\ }\bibfield  {title}
  {\enquote {\bibinfo {title} {Optical quantum super-resolution imaging and
  hypothesis testing},}\ }\href {\doibase 10.1038/s41467-022-32977-8}
  {\bibfield  {journal} {\bibinfo  {journal} {Nature Communications}\ }\textbf
  {\bibinfo {volume} {13}},\ \bibinfo {pages} {5373} (\bibinfo {year}
  {2022})}\BibitemShut {NoStop}%
\bibitem [{\citenamefont {Danilishin}\ \emph {et~al.}(2019)\citenamefont
  {Danilishin}, \citenamefont {Khalili},\ and\ \citenamefont
  {Miao}}]{danilishin19}%
  \BibitemOpen
  \bibfield  {author} {\bibinfo {author} {\bibfnamefont {Stefan~L.}\
  \bibnamefont {Danilishin}}, \bibinfo {author} {\bibfnamefont {Farid~Ya.}\
  \bibnamefont {Khalili}}, \ and\ \bibinfo {author} {\bibfnamefont {Haixing}\
  \bibnamefont {Miao}},\ }\bibfield  {title} {\enquote {\bibinfo {title}
  {Advanced quantum techniques for future gravitational-wave detectors},}\
  }\href {\doibase 10.1007/s41114-019-0018-y} {\bibfield  {journal} {\bibinfo
  {journal} {Living Reviews in Relativity}\ }\textbf {\bibinfo {volume} {22}},\
  \bibinfo {pages} {2} (\bibinfo {year} {2019})}\BibitemShut {NoStop}%
\bibitem [{\citenamefont {{M.~Tse \textit{et al.}}}(2019)}]{tse19}%
  \BibitemOpen
  \bibfield  {author} {\bibinfo {author} {\bibnamefont {{M.~Tse \textit{et
  al.}}}},\ }\bibfield  {title} {\enquote {\bibinfo {title} {Quantum-{Enhanced}
  {Advanced} {LIGO} {Detectors} in the {Era} of {Gravitational}-{Wave}
  {Astronomy}},}\ }\href {\doibase 10.1103/PhysRevLett.123.231107} {\bibfield
  {journal} {\bibinfo  {journal} {Physical Review Letters}\ }\textbf {\bibinfo
  {volume} {123}},\ \bibinfo {pages} {231107} (\bibinfo {year}
  {2019})}\BibitemShut {NoStop}%
\bibitem [{\citenamefont {Gefen}\ \emph {et~al.}(2019)\citenamefont {Gefen},
  \citenamefont {Rotem},\ and\ \citenamefont {Retzker}}]{gefen19}%
  \BibitemOpen
  \bibfield  {author} {\bibinfo {author} {\bibfnamefont {T.}~\bibnamefont
  {Gefen}}, \bibinfo {author} {\bibfnamefont {A.}~\bibnamefont {Rotem}}, \ and\
  \bibinfo {author} {\bibfnamefont {A.}~\bibnamefont {Retzker}},\ }\bibfield
  {title} {\enquote {\bibinfo {title} {Overcoming resolution limits with
  quantum sensing},}\ }\href {\doibase 10.1038/s41467-019-12817-y} {\bibfield
  {journal} {\bibinfo  {journal} {Nature Communications}\ }\textbf {\bibinfo
  {volume} {10}},\ \bibinfo {pages} {4992} (\bibinfo {year}
  {2019})}\BibitemShut {NoStop}%
\bibitem [{\citenamefont {Mouradian}\ \emph {et~al.}(2021)\citenamefont
  {Mouradian}, \citenamefont {Glikin}, \citenamefont {Megidish}, \citenamefont
  {Ellers},\ and\ \citenamefont {Haeffner}}]{mouradian21}%
  \BibitemOpen
  \bibfield  {author} {\bibinfo {author} {\bibfnamefont {Sara~L.}\ \bibnamefont
  {Mouradian}}, \bibinfo {author} {\bibfnamefont {Neil}\ \bibnamefont
  {Glikin}}, \bibinfo {author} {\bibfnamefont {Eli}\ \bibnamefont {Megidish}},
  \bibinfo {author} {\bibfnamefont {Kai-Isaak}\ \bibnamefont {Ellers}}, \ and\
  \bibinfo {author} {\bibfnamefont {Hartmut}\ \bibnamefont {Haeffner}},\
  }\bibfield  {title} {\enquote {\bibinfo {title} {Quantum sensing of
  intermittent stochastic signals},}\ }\href {\doibase
  10.1103/PhysRevA.103.032419} {\bibfield  {journal} {\bibinfo  {journal}
  {Physical Review A}\ }\textbf {\bibinfo {volume} {103}},\ \bibinfo {pages}
  {032419} (\bibinfo {year} {2021})}\BibitemShut {NoStop}%
\bibitem [{\citenamefont {G{\ifmmode\acute{o}\else{\'o}\fi}recki}\ \emph
  {et~al.}(2022)\citenamefont {G{\ifmmode\acute{o}\else{\'o}\fi}recki},
  \citenamefont {Riccardi},\ and\ \citenamefont {Maccone}}]{gorecki22}%
  \BibitemOpen
  \bibfield  {author} {\bibinfo {author} {\bibfnamefont {Wojciech}\
  \bibnamefont {G{\ifmmode\acute{o}\else{\'o}\fi}recki}}, \bibinfo {author}
  {\bibfnamefont {Alberto}\ \bibnamefont {Riccardi}}, \ and\ \bibinfo {author}
  {\bibfnamefont {Lorenzo}\ \bibnamefont {Maccone}},\ }\bibfield  {title}
  {\enquote {\bibinfo {title} {Quantum metrology of noisy spreading
  channels},}\ }\href {\doibase 10.48550/arXiv.2208.09386} {\bibfield
  {journal} {\bibinfo  {journal} {ArXiv e-prints}\ } (\bibinfo {year} {2022}),\
  10.48550/arXiv.2208.09386},\ \Eprint {http://arxiv.org/abs/2208.09386}
  {2208.09386} \BibitemShut {NoStop}%
\bibitem [{\citenamefont {Shi}\ and\ \citenamefont {Zhuang}(2022)}]{shi22}%
  \BibitemOpen
  \bibfield  {author} {\bibinfo {author} {\bibfnamefont {Haowei}\ \bibnamefont
  {Shi}}\ and\ \bibinfo {author} {\bibfnamefont {Quntao}\ \bibnamefont
  {Zhuang}},\ }\bibfield  {title} {\enquote {\bibinfo {title} {Ultimate
  precision limit of noise sensing and dark matter search},}\ }\href {\doibase
  10.48550/arXiv.2208.13712} {\bibfield  {journal} {\bibinfo  {journal} {ArXiv
  e-prints}\ } (\bibinfo {year} {2022}),\ 10.48550/arXiv.2208.13712},\ \Eprint
  {http://arxiv.org/abs/2208.13712} {2208.13712} \BibitemShut {NoStop}%
\bibitem [{\citenamefont {Tsang}\ \emph {et~al.}(2020)\citenamefont {Tsang},
  \citenamefont {Albarelli},\ and\ \citenamefont {Datta}}]{tsang20}%
  \BibitemOpen
  \bibfield  {author} {\bibinfo {author} {\bibfnamefont {Mankei}\ \bibnamefont
  {Tsang}}, \bibinfo {author} {\bibfnamefont {Francesco}\ \bibnamefont
  {Albarelli}}, \ and\ \bibinfo {author} {\bibfnamefont {Animesh}\ \bibnamefont
  {Datta}},\ }\bibfield  {title} {\enquote {\bibinfo {title} {Quantum
  {Semiparametric} {Estimation}},}\ }\href {\doibase
  10.1103/PhysRevX.10.031023} {\bibfield  {journal} {\bibinfo  {journal}
  {Physical Review X}\ }\textbf {\bibinfo {volume} {10}},\ \bibinfo {pages}
  {031023} (\bibinfo {year} {2020})}\BibitemShut {NoStop}%
\bibitem [{\citenamefont {Gross}\ and\ \citenamefont {Caves}(2020)}]{gross20}%
  \BibitemOpen
  \bibfield  {author} {\bibinfo {author} {\bibfnamefont {Jonathan~Arthur}\
  \bibnamefont {Gross}}\ and\ \bibinfo {author} {\bibfnamefont {Carlton~M.}\
  \bibnamefont {Caves}},\ }\bibfield  {title} {\enquote {\bibinfo {title} {One
  from many: {Estimating} a function of many parameters},}\ }\href {\doibase
  10.1088/1751-8121/abb9ed} {\bibfield  {journal} {\bibinfo  {journal} {Journal
  of Physics A: Mathematical and Theoretical}\ }\textbf {\bibinfo {volume}
  {54}},\ \bibinfo {pages} {014001} (\bibinfo {year} {2020})}\BibitemShut
  {NoStop}%
\bibitem [{\citenamefont {Kailath}(1967)}]{kailath67}%
  \BibitemOpen
  \bibfield  {author} {\bibinfo {author} {\bibfnamefont {Thomas}\ \bibnamefont
  {Kailath}},\ }\bibfield  {title} {\enquote {\bibinfo {title} {The divergence
  and {B}hattacharyya distance measures in signal selection},}\ }\href
  {\doibase 10.1109/TCOM.1967.1089532} {\bibfield  {journal} {\bibinfo
  {journal} {IEEE Transactions on Communication Technology}\ }\textbf {\bibinfo
  {volume} {15}},\ \bibinfo {pages} {52--60} (\bibinfo {year}
  {1967})}\BibitemShut {NoStop}%
\bibitem [{\citenamefont {Levy}(2008)}]{levy}%
  \BibitemOpen
  \bibfield  {author} {\bibinfo {author} {\bibfnamefont {Bernard~C.}\
  \bibnamefont {Levy}},\ }\href {\doibase 10.1007/978-0-387-76544-0} {\emph
  {\bibinfo {title} {Principles of Signal Detection and Parameter
  Estimation}}}\ (\bibinfo  {publisher} {Springer},\ \bibinfo {address} {New
  York},\ \bibinfo {year} {2008})\BibitemShut {NoStop}%
\end{thebibliography}%

\end{document}